\documentclass[twocolumn]{aastex631}

\newcommand\multimoon{\texttt{MultiMoon}}
\newcommand\jt{$J_2$}
\newcommand\ct{$C_{22}$}

\newcommand{\trackchange}[1]{#1}

\usepackage[utf8]{inputenc}
\usepackage{tipa}
\usepackage{combelow}
\usepackage{savesym}
\savesymbol{tablenum}
\usepackage{siunitx}
\restoresymbol{SIX}{tablenum}
\usepackage{breqn}
\usepackage{tabularx}
\usepackage{amsmath}

\shorttitle{Beyond Point Masses. III. Haumea}
\shortauthors{Proudfoot et al.}

\begin{document}

\title{Beyond Point Masses. III. Detecting Haumea's Nonspherical Gravitational Field}

\correspondingauthor{Benjamin Proudfoot}
\email{benp175@gmail.com}

\author[0000-0002-1788-870X]{Benjamin C.N. Proudfoot}
\affiliation{Brigham Young University, Department of Physics \& Astronomy, N283 ESC, Brigham Young University, Provo, UT 84602, USA}
\affiliation{Florida Space Institute, University of Central Florida, 12354 Research Parkway, Orlando, FL 32826, USA}
    
\author[0000-0003-1080-9770]{Darin A. Ragozzine}
\affiliation{Brigham Young University, Department of Physics \& Astronomy, N283 ESC, Brigham Young University, Provo, UT 84602, USA}

\author[0000-0001-6838-1530]{William Giforos}
\affiliation{Brigham Young University, Department of Physics \& Astronomy, N283 ESC, Brigham Young University, Provo, UT 84602, USA}
    
\author[0000-0002-8296-6540]{Will M. Grundy}
\affiliation{Lowell Observatory, 1400 W Mars Hill Rd, Flagstaff, AZ 86001, USA}
\affiliation{Northern Arizona University, Department of Astronomy \& Planetary Science, PO Box 6010, Flagstaff, AZ 86011, USA}

\author[0000-0003-2372-1364]{Mariah MacDonald}
\affiliation{The College of New Jersey, NJ, USA}

\author[0000-0001-5750-4953]{William J. Oldroyd}
\affiliation{Northern Arizona University, Department of Astronomy \& Planetary Science, PO Box 6010, Flagstaff, AZ 86011, USA}

\begin{abstract} 

The dwarf planet Haumea is one of the most compelling transneptunian objects (TNOs) to study, hosting two small, dynamically interacting satellites, a family of nearby spectrally unique objects, and a ring system. Haumea itself is extremely oblate due to its 3.9 hour rotation period. Understanding the orbits of Haumea's satellites, named Hi'iaka and Namaka, requires detailed modeling of both satellite-satellite gravitational interactions and satellite interactions with Haumea's nonspherical gravitational field (parameterized here as \jt). Understanding both of these effects allows for a detailed probe of the satellites' masses and Haumea's \jt{} and spin pole. Measuring Haumea's \jt{} provides information about Haumea's interior, possibly determining the extent of past differentation. In an effort to understand the Haumea system, we have performed detailed non-Keplerian orbit fitting of Haumea's satellites using a decade of new ultra-precise observations. Our fits detect Haumea's \jt{} and spin pole at $\gtrsim2.5\sigma$ confidence. Degeneracies present in the dynamics prevent us from precisely measuring Haumea's \jt{} with the current data, but future observations should enable a precise measurement. Our dynamically determined spin pole shows excellent agreement with past results, illustrating the strength of non-Keplerian orbit fitting. We also explore the spin-orbit dynamics of Haumea and its satellites, showing that axial precession of Hi'iaka may be detectable over decadal timescales. Finally, we present an ephemeris of the Haumea system over the coming decade, enabling high-quality observations of Haumea and its satellites for years to come. 

\end{abstract}

\section{Introduction}
\label{sec:intro}
Almost all of the largest transneptunian objects (TNOs) are known to host satellites \citep[e.g.][]{christy1978satellite,brown2005keck,brown2006satellites,brown2007satellites,noll2007binaries,parker2016discovery,kiss2017discovery}. These satellites are generally small relative to the system primary, and are thought to have formed by collisions \citep[][]{barr2016interpreting,arakawa2019early}. The current density of the transneptunian region is far too low to have formed so many satellites by collision \citep[][]{campo2012collisional,abedin2022ossos}, implying that these systems must not have formed in-situ. The emerging consensus is that large TNOs formed in a relatively massive primordial disk exterior to the giant planets, which was subsequently scattered by Neptune's outwards migration \citep[][]{nesvorny2018dynamical,gladman2021transneptunian}. The large TNOs we see today, which are on excited orbits, are the remnants of this primordial disk. \trackchange{By understanding how the satellites of large TNOs formed and evolve, we can probe the conditions of the early primordial disk where these systems formed.}

(136108) Haumea, the third most massive TNO known, is host to two satellites: Hi'iaka on a $\sim$50 day orbit and Namaka on a $\sim$20 day orbit \citep[][]{brown2005keck, brown2006satellites}. Haumea's shape, determined by both light curve observations \citep{rabinowitz2006photometric} and stellar occultations \citep{ortiz2017size}, is significantly nonspherical due to its 3.9 hour rotation period. Haumea and its satellites may have formed during a collision, which simultaneously spun up Haumea, created the satellites, and also formed Haumea's unique, icy collisional family \citep[][]{leinhardt2010formation,proudfoot2022formation}, but there remains some disagreement on these circumstances \citep[e.g.,][]{ortiz2012rotational,campo2016genesis,noviello2022let}. \trackchange{Connecting a formation model to all of the system's unique characteristics has been difficult despite many proposals} \citep[e.g.,][]{proudfoot2019modeling}. 

Our understanding of the complex nature of the Haumea system can be advanced by detailed study of the satellites' orbits. Study of the orbits potentially allows for a measurement of the masses of each component, as the two satellites strongly interact with each other \citep[][hereafter RB09]{rb09}. In addition to satellite-satellite interactions, perturbations from Haumea's nonspherical gravitational potential \trackchange{are also present}. Haumea's gravitational potential is determined both by its shape and internal density distribution. Since Haumea's shape is fairly well known \citep[due to observations of a stellar occultation, see][]{ortiz2017size}, measuring the gravitational harmonics of Haumea may allow us to constrain its internal density distribution. The internal density distributions of TNOs are almost completely unconstrained, although large TNOs are expected to be differentiated \citep{mckinnon2008structure,dunham2019haumea}. 

Haumea, in particular, has evidence for differentiation in the form of the collisional family. Haumea's family members are spectroscopically unique showing strong water ice features and few other major constituents (Souza Feliciano et al., in prep.). In combination with very high albedos \citep{elliot2010size,vilenius2014tnos}, it seems like the family members could be primarily composed of nearly pure water ice. This may suggest that the proto-Haumea was a differentiated ``ocean world'' and that Haumea family members are pieces of the water ice mantle. Thus studying present-day Haumea could give us unique insight into the interiors of ocean worlds. 

Haumea's gravitational potential can be described by a spherical harmonic expansion. To quadrupole order, the gravitational field of Haumea, $U$, with mass, $M$, at a distance $r$ can be written as:

\begin{dmath}\label{eqn:potential}
    U(r,\theta,\phi) = -\frac{GM}{r} \left[
    1 - J_2 \left( \frac{R}{r} \right) ^2 \left( \frac{3}{2}\sin^2\theta - \frac{1}{2} \right) 
    + C_{22} \left( \frac{R}{r} \right) ^2\cos^2\theta\sin2\phi + 
    \mathcal{O} \left( r^{-3} \right) \right]
\end{dmath}

\noindent where \jt{} and \ct{} are the second-order gravitational harmonic coefficients, $\theta$ is the body-fixed latitude angle, $\phi$ is the body-fixed longitude angle, and $R$ is an ``reference'' radius \citep[][]{1995geph.conf....1Y, scheeres2000evaluation}. In this work, we assume that $R$ is equivalent to volumetric radius. The coefficients \jt{} and \ct{} describe Haumea's shape and internal density structure. Taking the shape found by \cite{ortiz2017size} and assuming a homogeneous density, Haumea is expected to have $J_2 = 0.24$ and $C_{22} = 0.05$. However, when taking differentiation into account and using the \cite{dunham2019haumea} model of Haumea's interior, these harmonics would be $J_2 = 0.16$ and $C_{22} = 0.03$. While both of these models are simplified, they serve as a useful guide. 

In the original work that determined the orbits of Haumea's satellites (RB09), Haumea's nonspherical gravitational field was not clearly detected, although they were able to robustly detect satellite-satellite interactions. Subsequent follow up studies have also been unsuccessful in detecting the nonspherical field \citep[][]{gourgeot2016near}. However, with new ultra-precise HST observations from the past decade, another analysis of the satellites' orbits is in order. By leveraging these observations, as well as new computational techniques, we present a new, updated set of orbital fits to the Haumea system. We are able to detect the nonspherical gravitational potential of Haumea, constrain the masses of Haumea's satellites, and study the spin-orbit evolution of the system. 

The paper will be formatted as follows. In Section \ref{sec:obs}, we describe the observations used in our analysis. Then in Section \ref{sec:methods}, we describe our non-Keplerian orbital model and fitting procedure. Results of the fitting are presented in Section \ref{sec:results}, and discussed in Section \ref{sec:discussion}. We then conclude in Section \ref{sec:conclusions} and discuss future work. 

\section{Observations and Data Analysis}
\label{sec:obs}
The data we use in our orbit fitting comes from a variety of sources, but can be broadly broken into three separate datasets. The first dataset comes directly from RB09, which extracted satellite positions from Keck and HST observations. The second dataset consists of HST observations from HST Programs 12243 and 13873. The last dataset is made up of Keck observations from 2020-2022. For our orbit fitting, we combined the relative astrometry from each data set and simultaneously fit all the data. Our compiled data is presented in Table \ref{tab:observations}.

The published astrometry from RB09 was found to have a sign error in their listed RA offsets (in their Table 1). This error can be seen in their Figure 1 as RA decreases towards the east, opposite to convention. This mistake affected their orbital modeling, preventing them from correctly determining the orbital plane of the system, although the rest of their analysis is relatively unaffected. In our analysis, we use the RB09 data, although we correct the error and use the mirrored RA values. 

HST Programs 12243 and 13873 used HST's Wide Field Camera 3 (WFC3) to observe the Haumea system with a combined 13 orbits of coverage. Program 12243 imaged the system, over 10 consecutive HST orbits, in an attempt to observe a Haumea-Namaka mutual event. Program 13873 used 3 single-orbit visits to measure satellite relative astrometry to better constrain orbit models. Both of these programs took $\sim$30 individual exposures per orbit\trackchange{, with exposure times of 49 and 39 seconds for the programs respectively,} using the F350LP filter to maximize signal-to-noise. The images from these programs were analyzed using the same method used in RB09, although changes were made to fit the WFC3 data, replacing the older \trackchange{PSF models} used in previous studies. 

Keck observations used the laser guide star adaptive optics system (LGS AO) \citep{wizinowich2006} with the narrow camera of NIRC2\footnote{\url{https://www2.keck.hawaii.edu/inst/nirc2}}.  In the 2020 and 2021 observations, nearby field stars were used for tip-tilt correction, since they were brighter than Haumea.  All Keck observations were done in the infrared $H$ filter, covering wavelengths from $\sim$1.48 to 1.77 $\mu$m, with a series of dithered exposures for sky subtraction and to minimize the effect of bad pixels. \trackchange{120 second exposures were taken during the 2020 and 2022 observing runs, while 60 second exposures were used in 2021. Unfortunately, due to the short exposure time, Namaka was not visible in the 2021 images. During the 2022 observing run, unfavorable orbital phase also prevented detection of Namaka. Pairs of dithered images were flat-fielded and pairwise subtracted to remove sky background using practices common in TNO binary observations \citep[e.g.,][]{grundy2011five}.}

To extract relative astrometry of the satellites from the processed Keck data, we simultaneously fit 2-dimensional Gaussian PSFs to each visible object in individual processed images. While a Gaussian is a relatively poor approximation for the NIRC2 PSF, it is still able to measure the center of each PSF quite accurately. Relative detector positions were then converted to relative right ascension and declination assuming a mean plate scale of 9.952 mas/pixel and an orientation offset of 0.252$\degr$ \citep[][]{konopacky2010high,yelda2010improving,service2016new}. The median and standard deviation offsets of individual measurements are used for the astrometric offsets and error for each night, although we implemented a conservative noise floor of 10 milliarcseconds to account for unknown systematics. This method has been extensively used to extract relative astrometry from Keck NIRC2 images \citep[e.g.][]{fraser2010quaoar,grundy2015mutual}. 

As can be seen in Table \ref{tab:observations}, both satellites are not always detected at each epoch. In principle, non-detections can be used to help constrain the satellites' orbits, but in practice, given the already well-known orbits, they barely constrain the fits. Hence, during our orbit fitting process, we do not use non-detections in any way. 

\begin{deluxetable*}{ccccCCCCCCCC}
\tabletypesize{\footnotesize}
\tablewidth{\textwidth}
\tablecaption{Observed Astrometric Positions of Haumea's Satellites}
\tablehead{
Julian Date & Date & Telescope & Camera & \Delta x_N & \Delta y_N & \sigma_{\Delta x_N} & \sigma_{\Delta y_N} & \Delta x_H & \Delta y_H & \sigma_{\Delta x_H} & \sigma_{\Delta y_H} \\
 & & & & \textrm{('')} & \textrm{('')} & \textrm{('')} & \textrm{('')} & \textrm{('')} & \textrm{('')} & \textrm{('')} & \textrm{('')} 
}
\startdata
2453397.162 & 2005 Jan 26 & Keck & NIRC2 & \nodata & \nodata & \nodata & \nodata & -0.03506 & -0.63055 & 0.01394 & 0.01394 \\
2453431.009 & 2005 Mar 1 & Keck & NIRC2 & -0.00992 & 0.52801 & 0.02986 & 0.02986 & -0.29390 & -1.00626 & 0.02291 & 0.02291 \\
2453433.984 & 2005 Mar 4 & Keck & NIRC2 & \nodata & \nodata & \nodata & \nodata & -0.33974 & -1.26530 & 0.01992 & 0.01992 \\
2453518.816 & 2005 May 28 & Keck & NIRC2 & \nodata & \nodata & \nodata & \nodata & 0.06226 & 0.60575 & 0.00996 & 0.00996 \\
2453551.810 & 2005 Jun 30 & Keck & NIRC2 & 0.03988 & -0.65739 & 0.03978 & 0.03978 & 0.19727 & 0.52106 & 0.00498 & 0.00996 \\
2453746.525 & 2006 Jan 11 & HST & ACS/HRC & -0.04134 & -0.18746 & 0.00267 & 0.00267 & 0.20637 & 0.30013 & 0.00256 & 0.00256 \\
2453746.554 & 2006 Jan 11 & HST & ACS/HRC & -0.03867 & -0.19174 & 0.00267 & 0.00267 & 0.20832 & 0.30582 & 0.00257 & 0.00257 \\
2454138.287 & 2007 Feb 6 & HST & WFPC2 & 0.02627 & -0.57004 & 0.00702 & 0.00351 & 0.21088 & 0.22019 & 0.00252 & 0.00197 \\
2454138.304 & 2007 Feb 6 & HST & WFPC2 & 0.03107 & -0.56624 & 0.00210 & 0.00782 & 0.21132 & 0.22145 & 0.00095 & 0.00204 \\
2454138.351 & 2007 Feb 6 & HST & WFPC2 & 0.03009 & -0.55811 & 0.00527 & 0.00564 & 0.21515 & 0.23185 & 0.00301 & 0.00206 \\
2454138.368 & 2007 Feb 6 & HST & WFPC2 & 0.03133 & -0.56000 & 0.00482 & 0.00663 & 0.21402 & 0.23314 & 0.00192 & 0.00230 \\
2454138.418 & 2007 Feb 6 & HST & WFPC2 & 0.03134 & -0.54559 & 0.00385 & 0.00376 & 0.21705 & 0.24202 & 0.00103 & 0.00282 \\
2454138.435 & 2007 Feb 6 & HST & WFPC2 & 0.02791 & -0.54794 & 0.00571 & 0.00524 & 0.21449 & 0.24450 & 0.00323 & 0.00254 \\
2454138.484 & 2007 Feb 6 & HST & WFPC2 & 0.02972 & -0.53385 & 0.00797 & 0.01330 & 0.21818 & 0.25301 & 0.00153 & 0.00224 \\
2454138.501 & 2007 Feb 7 & HST & WFPC2 & 0.03226 & -0.53727 & 0.00531 & 0.00400 & 0.21807 & 0.25639 & 0.00310 & 0.00291 \\
2454138.551 & 2007 Feb 7 & HST & WFPC2 & 0.03429 & -0.53079 & 0.00497 & 0.00582 & 0.22173 & 0.26308 & 0.00146 & 0.00230 \\
2454138.567 & 2007 Feb 7 & HST & WFPC2 & 0.03576 & -0.52712 & 0.00270 & 0.00479 & 0.21978 & 0.26791 & 0.00202 & 0.00226 \\
2454469.653 & 2008 Jan 4 & HST & WFPC2 & 0.02399 & -0.28555 & 0.00670 & 0.00831 & -0.23786 & -1.27383 & 0.00404 & 0.00824 \\
2454552.897 & 2008 Mar 27 & Keck & NIRC2 & \nodata & \nodata & \nodata & \nodata & -0.19974 & -0.10941 & 0.00930 & 0.00956 \\
2454556.929 & 2008 Mar 31 & Keck & NIRC2 & -0.00439 & -0.76848 & 0.01239 & 0.01280 & -0.32988 & -0.77111 & 0.00455 & 0.00557 \\
2454556.948 & 2008 Mar 31 & Keck & NIRC2 & -0.01363 & -0.76500 & 0.01976 & 0.01252 & -0.33367 & -0.77427 & 0.00890 & 0.00753 \\
2454556.964 & 2008 Mar 31 & Keck & NIRC2 & -0.00576 & -0.77375 & 0.01212 & 0.01283 & -0.33267 & -0.77874 & 0.00676 & 0.00485 \\
2454557.004 & 2008 Mar 31 & Keck & NIRC2 & -0.00854 & -0.77313 & 0.01199 & 0.00897 & -0.33543 & -0.78372 & 0.00404 & 0.00592 \\
2454557.020 & 2008 Mar 31 & Keck & NIRC2 & -0.00075 & -0.76974 & 0.00907 & 0.01015 & -0.33491 & -0.78368 & 0.00374 & 0.00473 \\
2454557.039 & 2008 Mar 31 & Keck & NIRC2 & -0.00988 & -0.77084 & 0.01793 & 0.01543 & -0.33712 & -0.78464 & 0.00740 & 0.00936 \\
2454557.058 & 2008 Mar 31 & Keck & NIRC2 & -0.01533 & -0.76117 & 0.00765 & 0.01571 & -0.33549 & -0.78692 & 0.00868 & 0.00852 \\
2454557.074 & 2008 Mar 31 & Keck & NIRC2 & -0.00645 & -0.76297 & 0.01639 & 0.01390 & -0.33128 & -0.78867 & 0.01431 & 0.01411 \\
2454557.091 & 2008 Mar 31 & Keck & NIRC2 & -0.00708 & -0.76986 & 0.01532 & 0.00787 & -0.33687 & -0.79462 & 0.00803 & 0.00717 \\
2454593.726 & 2008 May 7 & HST & NICMOS & -0.00243 & -0.75878 & 0.00576 & 0.00761 & 0.18297 & 1.08994 & 0.00354 & 0.00425 \\
2454600.192 & 2008 May 13 & HST & WFPC2 & 0.02325 & 0.19934 & 0.00480 & 0.01161 & -0.10847 & 0.17074 & 0.00508 & 0.00427 \\
2454601.990 & 2008 May 15 & HST & WFPC2 & 0.02293 & 0.50217 & 0.00618 & 0.00614 & -0.18374 & -0.13041 & 0.00729 & 0.00504 \\
2454603.788 & 2008 May 17 & HST & WFPC2 & 0.01174 & 0.59613 & 0.00366 & 0.00485 & -0.24918 & -0.43962 & 0.00207 & 0.00574 \\
2454605.788 & 2008 May 19 & HST & WFPC2 & -0.00006 & 0.29915 & 0.00425 & 0.00613 & -0.29818 & -0.75412 & 0.00467 & 0.00966 \\
2455375.655 & 2010 Jun 28 & HST & WFC3 & 0.00735 & 0.19620 & 0.00168 & 0.00161 & \nodata & \nodata & \nodata & \nodata \\
2455375.661 & 2010 Jun 28 & HST & WFC3 & \nodata & \nodata & \nodata & \nodata & 0.26874 & 1.22502 & 0.00159 & 0.00154 \\
2455375.673 & 2010 Jun 28 & HST & WFC3 & 0.00766 & 0.18829 & 0.00326 & 0.00336 & \nodata & \nodata & \nodata & \nodata \\
2455375.719 & 2010 Jun 28 & HST & WFC3 & 0.00729 & 0.18426 & 0.00202 & 0.00778 & \nodata & \nodata & \nodata & \nodata \\
2455375.727 & 2010 Jun 28 & HST & WFC3 & \nodata & \nodata & \nodata & \nodata & 0.26632 & 1.22294 & 0.00126 & 0.00164 \\
2455375.737 & 2010 Jun 28 & HST & WFC3 & 0.00612 & 0.17861 & 0.00170 & 0.00252 & \nodata & \nodata & \nodata & \nodata \\
2455375.786 & 2010 Jun 28 & HST & WFC3 & 0.00926 & 0.16304 & 0.00144 & 0.00274 & \nodata & \nodata & \nodata & \nodata \\
2455375.793 & 2010 Jun 28 & HST & WFC3 & \nodata & \nodata & \nodata & \nodata & 0.26374 & 1.22053 & 0.00138 & 0.00193 \\
2455375.859 & 2010 Jun 28 & HST & WFC3 & \nodata & \nodata & \nodata & \nodata & 0.26187 & 1.21840 & 0.00131 & 0.00182 \\
2455375.928 & 2010 Jun 28 & HST & WFC3 & \nodata & \nodata & \nodata & \nodata & 0.25945 & 1.21625 & 0.00150 & 0.00175 \\
2455375.993 & 2010 Jun 28 & HST & WFC3 & \nodata & \nodata & \nodata & \nodata & 0.25813 & 1.21560 & 0.00137 & 0.00189 \\
2455376.058 & 2010 Jun 28 & HST & WFC3 & \nodata & \nodata & \nodata & \nodata & 0.25598 & 1.21306 & 0.00165 & 0.00136 \\
2456995.589 & 2014 Dec 4 & HST & WFC3 & -0.04910 & -0.34609 & 0.00200 & 0.00222 & 0.17725 & 1.13669 & 0.00200 & 0.00200 \\
2457155.338 & 2015 May 12 & HST & WFC3 & -0.09964 & -0.45547 & 0.00315 & 0.00433 & -0.44571 & -0.69806 & 0.00454 & 0.00568 \\
2457203.995 & 2015 Jun 30 & HST & WFC3 & 0.14931 & 0.69611 & 0.00200 & 0.00200 & -0.42272 & -0.63347 & 0.00200 & 0.00200 \\
2458885.090 & 2020 Feb 5 & Keck & NIRC2 & 0.21330 & 0.29118 & 0.01000 & 0.01000 & -0.03064 & -1.15403 & 0.01000 & 0.01000 \\
2459272.041 & 2021 Feb 26 & Keck & NIRC2 & \nodata & \nodata & \nodata & \nodata & -0.37255 & -1.36839 & 0.01000& 0.01000 \\
2459598.127 & 2022 Jan 18 & Keck & NIRC2 & \nodata & \nodata & \nodata & \nodata & -0.13988 & 0.80436 & 0.01000 & 0.01000 \\
\enddata
\tablecomments{The relative right ascension and declination positions of Haumea's satellites, Hi'iaka (H) and Namaka (N). At some epochs, Hi'iaka or Namaka were not visible in the images, for a variety of reasons. For these entries, no data is listed and our orbit fits were not constrained by their non-detection. Data from before 2010 are taken from \citet{rb09}, although we correct their sign error in the $\Delta x$ columns.}
\label{tab:observations}
\end{deluxetable*}

\section{Methods}
\label{sec:methods}
For our orbit fitting, we use \multimoon, a state-of-the-art orbit fitter designed for use with TNOs (Ragozzine et al., submitted). \multimoon{} is built around an n-quadrupole integrator that can simulate the \trackchange{gravitational and rotational dynamics of an arbitrary number of tri-axial ellipsoids to quarupole order}. Internally, it uses \texttt{emcee} \citep[][]{foreman2013emcee,foreman2019emcee}, a popular Markov chain Monte Carlo (MCMC) ensemble sampler, allowing us to treat the orbit fit as a Bayesian inference problem. In its simplest form, \multimoon{} uses a least squares method for evaluating the goodness-of-fit of a given orbital model. \trackchange{It can also accommodate a more complicated goodness-of-fit metric which we describe later in this paper.}

\trackchange{In our fits, we only consider the \jt{} since it is by far the most dominant harmonic.} The other second order harmonic \ct{}--which is related to the prolateness of Haumea--is relevant for understanding dynamics of orbits around Haumea, but only within a few times the corotation radius \citep{proudfoot2021prolate}. Even within this range, \ct{} averages out over many orbits except when close to a spin-orbit resonance (SOR). Haumea also certainly has substantial higher-order harmonics (most notably $J_4$), but their effect is small due to their $r^{-5}$ distance dependence. As further justification of this assumption, we can analytically estimate the precession induced by Haumea's $J_4$, and find it is only $\sim$0.1\% the strength of \jt{} precession for Namaka, and even smaller for Hi'iaka. Thus we believe that our simple model of Haumea's gravitational potential is sufficient to describe the dynamics taking place in the Haumea system. 

For the orbit fits presented here, we only model the gravitational harmonics of Haumea ignoring the (presumably) nonspherical shapes of Hi'iaka and Namaka. We revisit this assumption later in the paper. \trackchange{Our baseline orbit model has 18 free parameters including the mass, \jt, and 2 spin pole direction angles of Haumea, in addition to the masses and 6 orbital elements of each satellite.} Our model also requires the input of Haumea's rotational period to correctly model any axial precession which the satellites may cause. Although this value could, in principle, be a free parameter in the model, it is known with high precision and has very little influence on the orbital dynamics of the system. Hence we opt to use a fixed value of 3.915 hours \citep[][]{rabinowitz2006photometric}. 

To account for possible systematics arising from the use of a variety of data sets, we have implemented a sophisticated likelihood function within \multimoon{}. This likelihood function is adapted from the outlier pruning methods presented in \citet{hogg2010data}. Since we, \textit{a priori}, do not describe the systematic errors that may arise in the fitting process, we use an extremely flexible framework. Our likelihood model is a mixture model that combines two least-squares terms. The first is a common least-squares likelihood model, the standard technique for orbit fitting. This term is combined with another least-squares model with an additional error term. The error term, which we call $\sigma_{sys}$, is combined with the measured uncertainties of our observations in quadrature. Also included is a normalization factor ($f_{sys}$) describing the fraction of data displaying systematic errors, which also acts as a penalty for exclusion of data. The entire likelihood function can be written:

\begin{dmath}
    \mathcal{L} = \prod_{i=1}^{N} \Biggr[
    \left( \frac{1 - f_{sys}}{\sqrt{2\pi \sigma_i^2}} \right) 
    \exp{ \left(-\frac{(y_i - y_{i,m})^2}{2\sigma_i^2} \right)} +                 
    \left( \frac{f_{sys}}{\sqrt{2\pi (\sigma_i^2 + \sigma_{sys}^2)}} \right) 
    \exp{ \left(-\frac{(y_i - y_{i,m})^2}{2(\sigma_i^2 + \sigma_{sys}^2)} \right)} \Biggr]
\end{dmath}

\noindent where $y_i$ and $\sigma_i$ are the $N$ observations and uncertainties, and $y_{i,m}$ is the model. Technically, $y_{i,m}$ and $\sigma_i$ are vectors where each have two dimensions ($\Delta \alpha \cos{\delta}$, $\Delta \delta$) and there is an implied summation over both of these dimensions. For brevity, however, we exclude this summation, although it is implemented in its full form internally. If there are significant outliers in the data, this prescription downweights them relative to the typical least-squares assumptions and thus qualifies as a ``robust'' (to outliers) statistical method. In this sense, it operates similar to an automated sigma-clipping technique. It also allows for the expansion of systematic uncertainties when the quoted statistical uncertainties are too small to explain the scatter in residuals relative to the model.

The factors $(1-f_{sys})$ and $f_{sys}$ are critical for normalizing the two likelihood models and provide an implicit prior that penalizes overestimation of systematic effects. However, when $\sigma_{sys} \ll \sigma_{i}$ (i.e. there are no systematic errors present in the data), $f_{sys}$ is not well defined as both likelihood functions asymptotically approach one another. To prevent this degeneracy from becoming problematic, we implement a prior forcing $\sigma_{sys} \geq$ 1 milliarcsecond. This likelihood model adds an additional two free parameters to our model ($\sigma_{sys}$, $f_{sys}$). However, instead of fitting $\sigma_{sys}$, we opt to fit $\log_{10}{\sigma_{sys}}$, allowing the MCMC algorithm to more easily explore a greater range of values. This ``robust'' likelihood model has now been implemented into \multimoon{} and is publicly available on GitHub\footnote{\url{github.com/dragozzine/multimoon}}. We have extensively validated this likelihood model using synthetically produced data sets that have large systematics applied. We find that when using this model, \multimoon{} can recover the original \trackchange{orbital} parameters even when systematic uncertainties of 10s of milliarcseconds are applied to $\sim$50\% of the data set. 

During our data fitting process, we found that large systematics were present when combining both the Keck and HST data that necessitated the use of this robust likelihood model. Unfortunately, our model could not resolve these issues and unusual systematics remained unaccounted for. To remain as conservative as possible, we elected to complete an orbit fit using the HST data only with a standard least-squares likelihood model. We discuss the drawbacks of the HST+Keck fit further in Section \ref{sec:results} and \ref{sec:discussion}.

\begin{figure*}
    \includegraphics[width=\textwidth]{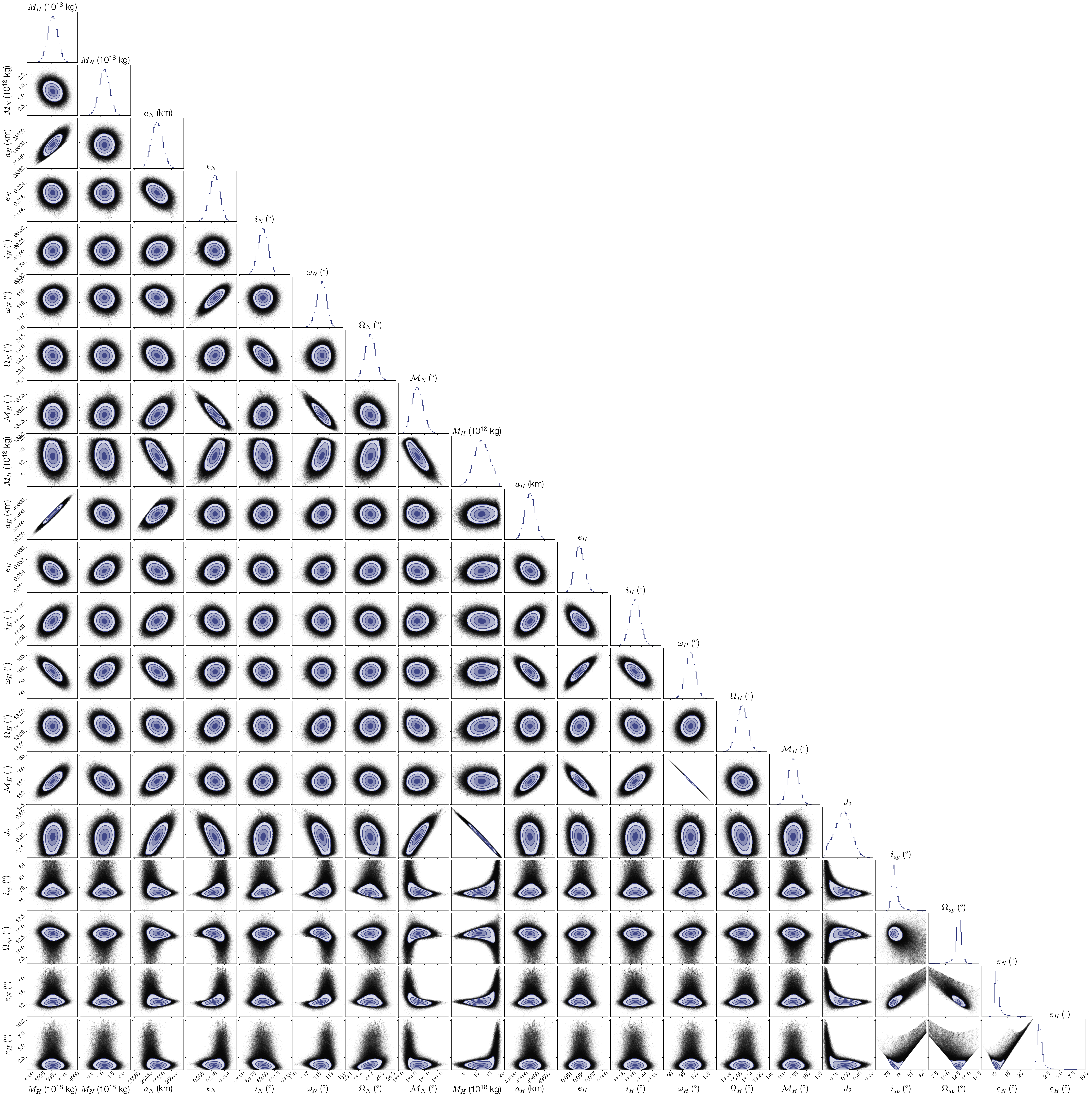}
    \caption{A corner plot for the HST only orbit fit to the Haumea system. We include all 18 fitted parameters along with 2 derived parameters. To facilitate easy interpretation, we list \jt{} rather than $J_2R^2$ by taking the volumetric radius $R$ from the occultation derived shape model \citep{ortiz2017size}. We also show the inclinations of each satellite with respect to Haumea's equator in the last two columns. Along the top of each column is the marginal posterior distribution for each parameter in our fit. Below the marginal distributions are the 2-dimensional joint posterior distributions for every pair of parameters. Contours correspond to 1, 2, and 3$\sigma$ levels. Small black points mark individual samples from our MCMC chains. The best fit parameter set in our MCMC chains corresponds to a $\chi^2$ of 99.1 with 90 degrees of freedom. Of particular note is the strong exclusion of $J_2 = 0$ in the marginal posterior for Haumea's \jt, alongside strong correlations between \jt{} and a variety of other parameters. }
    \label{fig:corner}
\end{figure*}

As part of the Bayesian framework \multimoon{} uses, we set priors for all parameters to be uninformative (except for $\sigma_{sys}$ as discussed above), allowing the data to constrain the posterior distribution. However, in our HST only fit, we set uniform priors on the spin pole direction of Haumea to prevent walkers from getting stuck in a lower dimensional subspace. The priors were chosen to bracket the best region of likelihood space within $\sim10\degr$, as identified in preliminary fits. After the fit was completed, we confirmed that this prior did not significantly prevent walkers from exploring favorable parts of likelihood space.

We drew initial walker positions from Gaussian distributions centered near the location of highest likelihood that was identified in preliminary runs. These preliminary runs were conducted to broadly search parameter space and used very broad initial guesses, allowing for a rigorous search of the 18-dimensional parameter space (20-dimensional for the HST+Keck fits). Our preliminary fits showed no signs of other likelihood maxima with acceptable fit quality. Our baseline orbit fits used 960 walkers in the MCMC ensemble, which were run for 5000 burn-in steps. We then pruned underperforming walkers, replacing them with random linear combinations of highly performing walkers, after which the ensemble was run for 1000 more burn-in steps. The ensemble was then run for 20000 steps to sample the posterior distribution. We confirmed that the resulting chains were converged by visual inspection of walker trace plots and marginal parameter-likelihood plots.

\begin{deluxetable*}{lcLL}
\tablecaption{Non-Keplerian Orbit Solutions for Haumea's Satellites}
\tablehead{
Parameter &  & \text{HST only fit} & \text{HST+Keck fit}
}
\startdata
Fitted parameters & & & \\
\qquad Mass, Haumea ($10^{18}$ kg) & $M_{P}$ & $3952.44^{+11.09}_{-11.03}$ & $3952.62^{+9.33}_{-9.09}$ \\
\qquad Mass, Namaka ($10^{18}$ kg) & $M_N$ & $1.18^{+0.25}_{-0.25}$ & $1.1^{+0.17}_{-0.18}$ \\
\qquad Semi-major axis, Namaka (km) & $a_N$ & $25506^{+36}_{-36}$ & $25548^{+27}_{-28}$ \\
\qquad Eccentricity, Namaka & $e_N$ & $0.2179^{+0.0032}_{-0.0033}$ & $0.2137^{+0.0042}_{-0.0043}$ \\
\qquad Inclination, Namaka ($\degr$) & $i_N$ & $69.005^{+0.108}_{-0.107}$ & $69.048^{+0.103}_{-0.103}$ \\
\qquad Argument of periapse, Namaka ($\degr$) & $\omega_N$ & $118.35^{+0.39}_{-0.42}$ & $117.82^{+0.58}_{-0.60}$ \\
\qquad Longitude of the ascending node, Namaka ($\degr$) & $\Omega_N$ & $23.725^{+0.149}_{-0.15}$ & $23.606^{+0.162}_{-0.154}$ \\
\qquad Mean anomaly at epoch, Namaka($\degr$) & $\mathcal{M}_N$ & $185.19^{+0.69}_{-0.65}$ & $186.32^{+0.73}_{-0.70}$ \\
\qquad Mass, Hi'iaka ($10^{18}$ kg) & $M_H$ & $12.13^{+3.22}_{-3.11}$ & $6.65^{+1.67}_{-1.52}$ \\
\qquad Semi-major axis, Hi'iaka (km) & $a_H$ & $49371^{+45}_{-45}$ & $49352^{+37}_{-35}$ \\
\qquad Eccentricity, Hi'iaka & $e_H$ & $0.0542^{+0.0012}_{-0.0012}$ & $0.0545^{+0.0009}_{-0.0009}$ \\
\qquad Inclination, Hi'iaka ($\degr$) & $i_H$ & $77.394^{+0.038}_{-0.038}$ & $77.376^{+0.035}_{-0.035}$ \\
\qquad Argument of periapse, Hi'iaka ($\degr$) & $\omega_H$ & $98.34^{+2.02}_{-2.06}$ & $99.05^{+1.48}_{-1.49}$ \\
\qquad Longitude of the ascending node, Hi'iaka ($\degr$) & $\Omega_H$ & $13.11^{+0.030}_{-0.031}$ & $13.071^{+0.030}_{-0.029}$ \\
\qquad Mean anomaly at epoch, Hi'iaka ($\degr$) & $\mathcal{M}_H$ & $154.53^{+2.05}_{-2.00}$ & $153.88^{+1.48}_{-1.47}$ \\
\qquad Second zonal gravitational harmonic & $J_2$ & $0.262^{+0.103}_{-0.112}$ & $0.431^{+0.046}_{-0.051}$ \\
\qquad Rotation axis obliquity ($\degr$) & $i_{sp}$ & $76.83^{+1.03}_{-0.59}$ & $75.32^{+0.68}_{-0.59}$ \\
\qquad Rotation axis precession ($\degr$) & $\Omega_{sp}$ & $13.1^{+0.65}_{-0.75}$ & $13.4^{+0.57}_{-0.52}$ \\
\qquad Systematic error fraction & $f_{sys}$ & \nodata & $0.122^{+0.115}_{-0.065}$ \\
\qquad Systematic error uncertainty & $\log_{10}(\sigma_{sys}/1'')$ & \nodata & $-2.085^{+0.169}_{-0.201}$ \\
Derived parameters & & & \\
\qquad Inclination w.r.t. Haumea's equator, Namaka ($\degr$) & $\varepsilon_N$ & $12.79^{+1.01}_{-0.58}$ & $11.56^{+0.70}_{-0.65}$ \\
\qquad Inclination w.r.t. Haumea's equator, Hi'iaka ($\degr$) & $\varepsilon_H$ & $1.01^{+0.66}_{-0.47}$ & $2.13^{+0.63}_{-0.68}$ \\
\qquad Haumea pole right ascension ($\degr$) & $\alpha_{p}$ & $282.9^{+0.6}_{-0.7}$ & $283.1^{+0.5}_{-0.5}$ \\
\qquad Haumea pole declination ($\degr$) & $\delta_{p}$ & $-9.7^{+0.6}_{-1.0}$ & $-8.1^{+0.6}_{-0.7}$ \\
\qquad Orbit pole right ascension, Namaka ($\degr$) & $\alpha_{N}$ & $292.1^{+0.1}_{-0.1}$ & $292.0^{+0.1}_{-0.1}$ \\
\qquad Orbit pole declination, Namaka ($\degr$) & $\delta_{N}$ & $-0.6^{+0.1}_{-0.1}$ & $-0.7^{+0.1}_{-0.1}$ \\
\qquad Orbit pole right ascension, Hi'iaka ($\degr$) & $\alpha_{H}$ & $283.00^{+0.03}_{-0.03}$ & $282.96^{+0.03}_{-0.03}$ \\
\qquad Orbit pole declination, Hi'iaka ($\degr$) & $\delta_{H}$ & $-10.24^{+0.04}_{-0.04}$ & $-10.23^{+0.04}_{-0.04}$ \\ 
\hline
\enddata
\tablecomments{Reported values represent the median value taken from the posterior distribution, while the stated uncertainties represent the 16th and 84th percentiles. All fitted angles are relative to the J2000 ecliptic plane on Haumea-centric JD 2454615.0 (2008 May 28 12:00 UT), chosen to match the epoch used in \citet{rb09}. Assumed c-axis for Haumea is 537 km \citep{dunham2019haumea} and spin period is 3.915 hours \citep{rabinowitz2006photometric}, however, altering these values produces no meaningful change to the fit. To transform to $J_2$ from only the more physically meaningful $J_2R^2$, we use a volumetric radius of 798 km \citep{ortiz2017size}. }
\label{tab:fits}
\end{deluxetable*}

\section{Results}
\label{sec:results}
When comparing our different orbit fits, we find that there is strong disagreement between the two datasets (HST+Keck and HST only). When fitting to the combined dataset, we find that the most recent Keck observations of the system are at odds with the 2014-2015 HST observations. Using our robust likelihood model and the combined HST+Keck dataset, we find that our best fit is $\sim$10$\sigma$ inconsistent with the 2014 HST observation of Namaka. This inconsistency was attributable to the data rather than the model, as shown by fits both with and without our robust likelihood model. Our robust likelihood model parameters indicated that approximately 10\% of the data had uncertainties underestimated by $\sim$10 milliarcseconds. Fits without the robust likelihood model were extremely similar to those with it, except the fit quality was much worse. When closely examined, no obvious problems were seen in the Keck or HST images, and no difficulties arose during our analysis of the images. To examine whether our image analysis techniques were to blame, we attempted to extract astrometry from the images with a variety of techniques (e.g., Gaussian PSF fitting, WFC3 model PSF fitting, etc.), all of which yielded similar results. 

In addition to the internal inconsistency, the HST-Keck combined fit also produced a measurement for Haumea's \jt{} that was too high to be compatible with other observations of the Haumea system (see Section \ref{sec:discussion} for more discussion). In comparison, the HST only fit showed no such issues. When the orbit fits are compared, very little changes between the models with the exception of Hi'iaka's mass, Haumea's \jt{}, and Haumea's spin pole direction. As unknown systematic errors are affecting our orbit fit, we choose to proceed by eliminating possible sources of these systematic errors. Since HST's PSF is extremely stable and has been extensively cross-calibrated across instruments, we adopt the HST only orbit fits for the purpose of this work. This choice results in larger uncertainties within the model, but allows us to be more confident that our results are not affected by systematics. Although we adopt the HST only fit, we still discuss the implications of our combined orbit fit in Section \ref{sec:discussion}, as well as reporting its results in Table \ref{tab:fits}.

\begin{figure}
    \includegraphics[width=\linewidth]{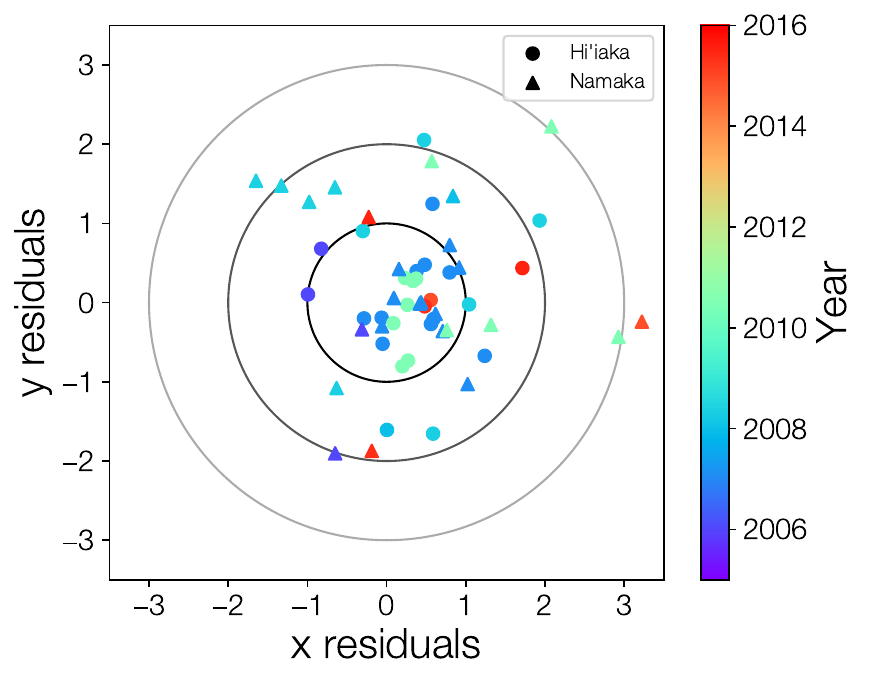}
    \caption{The normalized residuals of the best parameter set from our HST-only (non-robust) orbit fit. $x$ and $y$ correspond to ecliptic longitude and latitude, which is the primary coordinate system used in \multimoon. Hi'iaka residuals are shown as circles and Namaka with triangles. The color of each point corresponds to the observation date. The circles correspond to 1, 2, and 3$\sigma$ error contours. As reported in the text, this fit corresponded to a reduced $\chi^2 \sim 1.10$. We find that although the fit quality is worse than would be desired, the p-value of associated with the $\chi^2$ statistic is 0.23, indicating an acceptable fit. }
    \label{fig:target}
\end{figure}

The results presented here are our most refined orbital fits. Including our preliminary analysis, exploratory fits, and fits using different likelihood models, our nominal orbit model is the result of well over $10^9$ individual orbit integrations. We show the HST only orbit model in it entirety as a corner plot \citep{foreman2016corner} in Figure \ref{fig:corner}. Each column in the corner plot displays the marginal posterior distribution for each parameter as a histogram (at the top) and 2-dimensional joint posterior distribution as a contour plot. In addition, we display the marginal posteriors for both fits in Table \ref{tab:fits}. Both Figure \ref{fig:corner} and Table \ref{tab:fits} also contain several derived parameters, parameters that are functions of our fitted parameters. To display the fit quality, we show the residuals of the best fit parameter set in Figure \ref{fig:target}, alongside 1, 2, and 3 $\sigma$ error contours. \trackchange{This best fit parameter set is only one realization of our posterior distribution, but it illustrates the quality of our fit.}

\begin{figure}
    \includegraphics[width=\linewidth]{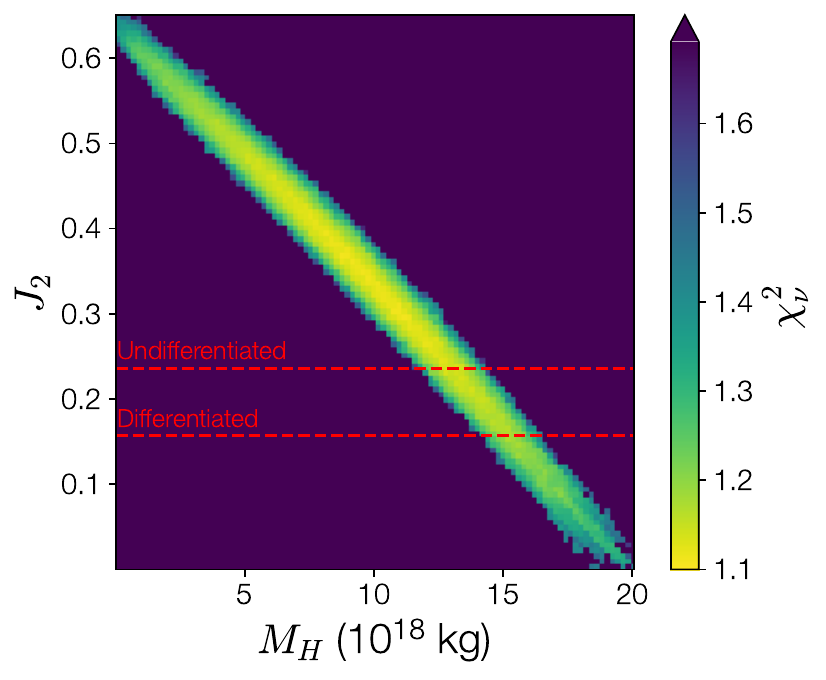}
    \caption{The joint Haumea \jt--Hi'iaka mass posterior distribution. Instead of displaying the density of sampled points as in Figure \ref{fig:corner}, we show the maximum fit quality (as measured by reduced $\chi^2$) in a small bin. Bins without any sampled points, indicating extremely poor quality fits, were set to the minimum bin value, although the true value is likely much worse. We find that the best fit with $\chi^2_{\nu} \sim 1.1$ has a p-value of 0.23, meaning there is a 23\% chance that random chance would produce a worse fit. A $\chi^2_{\nu} \sim 1.25$ corresponds to a p-value of 0.05. Dashed red lines show the expected \jt{} values from different internal density models. The undifferentiated model assumes a homogeneous interior along with the occultation derived shape model \citep{ortiz2017size}. The differentiated model is a two layer model proposed by \citet{dunham2019haumea}. The posterior shows that both models are consistent with the data, although the differentiated model is slightly disfavored. }
    \label{fig:degneracy}
\end{figure}

One of the outstanding features of our orbit fit is our detection of Haumea's \jt. When assuming the volumetric radius derived from stellar occultation measurements \citep[798 km, ][]{ortiz2017size}, we find $J_2 = 0.262$. However, our orbit fit shows that Haumea's \jt{} and Hi'iaka's mass are highly degenerate with one another. In Figure \ref{fig:degneracy}, we show, in detail, the degeneracy between these parameters as a function of reduction in fit quality. It is clear that a large range of values for these two parameters are acceptable, with nearly no reduction in fit quality. In our HST+Keck orbit fit, we find that Haumea's \jt{} has much lower uncertainties, but is unexpectedly high, $J_2 = 0.431$. Although probably attributable to unidentified systematic errors in our dataset, we will discuss possible causes/interpretations of this unusual measurement in Section \ref{sec:discussion}. Our detection of Haumea's \jt{} is significant in both orbit fits. The HST only fit detects \jt{} at $\sim$2.5$\sigma$ confidence, while the HST+Keck fit detects it at $>5\sigma$ confidence. Alongside our detection and measurement of Haumea's \jt, we also provide a measurement of Haumea's rotation pole. We find that Haumea's pole (or more precisely, the pole of Haumea's gravitational quadrupole) points toward $(\alpha_{p},\delta_{p}) = (282.9^{+0.6}_{-0.7}\degr,-9.7^{+0.6}_{-1.0}\degr)$, very close to the occultation derived rotation pole \trackchange{of $(\alpha_{p},\delta_{p}) = (285.1\pm0.5\degr,-10.6\pm1.2\degr)$} \citep[][]{ortiz2017size}. 

In our orbit fit, we are able to significantly detect the masses of both satellites at significance $>3\sigma$. RB09 previously detected Namaka and Hi'iaka's masses, but only with 1.2$\sigma$ confidence for Namaka. While our fit strongly detects both, the uncertainty on Hi'iaka's mass is substantial due to its degeneracy with Haumea's \jt. Alongside mass measurements, we are also able to constrain the satellites' inclinations with respect to Haumea's equator. We find inclinations of 12.8$^{+0.8}_{-0.6} \degr$ and 1.0$^{+0.6}_{-0.5} \degr$, for Namaka and Hi'iaka, respectively. We also measure the satellites' mutual inclination of 13.2$^{+0.2}_{-0.2} \degr$.

Our orbit fits are significantly different from past orbit fits \citep{rb09,gourgeot2016near}. While this difference is expected since we include more dynamical effects (e.g. including \jt), some important differences are still present. Most notable is the change in orbit angles, which stems from RB09's incorrectly tabulated astrometry, allowing for close agreement with the orbit planes found in \citet{gourgeot2016near}. We find a lower eccentricity for Namaka ($0.2179^{+0.0032}_{-0.0033}$) compared to RB09 ($0.249 \pm 0.015$ using the same epoch), also presumably due to their incorrect astrometry. Another notable difference is the change in Hi'iaka's mass ($12.13^{+3.22}_{-3.11} \times 10^{18}$ kg) when compared to RB09 ($17.9\pm1.1 \times 10^{18}$ kg), due to our inclusion of Haumea's \jt. Our preliminary fits showed that our orbit model, when evaluated with a small \jt{} approximately reproduces RB09's measurement of Hi'iaka's mass. 

When compared to the orbit model presented in \citet{gourgeot2016near}, we find quite large differences in orbital parameters especially in the fit for Namaka's orbit. This is unsurprising since their orbit model was a pure Keplerian orbit fit, neglecting both Haumea's \jt{} and satellite-satellite interactions. Their analysis claimed that there was no signature of non-Keplerian effects caused by Haumea's \jt{} in the system, although they use a much shorter span of data than our analysis. We find that non-Keplerian effects from both satellite-satellite interactions and Haumea's \jt{} are strongly detected, however it remains uncertain how strong each effect is. 

\section{Discussion}
\label{sec:discussion}

\subsection{Haumea's Large $J_2$}
\label{sec:dis_j2}
Assuming a homogeneous density structure and using the equations found in \citet{1995geph.conf....1Y}, the occultation derived shape implies a \jt{} of 0.24 \citep{ortiz2017size}. Allowing for differentiation decreases \jt{} significantly. The model for a two-layer differentiated Haumea presented in \citet{dunham2019haumea}, gives an overall \jt{} of $\sim$0.16. Our model fitting to all available data (HST+Keck) is 3$\sigma$ inconsistent with both of these models. However, the fit with only HST data is consistent with both, encouraging us to explore possible reasons Haumea's \jt{} may be higher. 

One possible reason could be the gravitational contributions from Haumea's ring. Assuming a circular ring, the following expression can be derived for the \jt{} contribution of a ring:

\begin{align}
    J_{2}R^2 \approx \frac{1}{2} \frac{M_{r}}{M_P} r^2
\label{eqn:ringj2}
\end{align}
\noindent where $M_r$ and $M_P$ are the masses of the ring and Haumea, respectively, and $r$ is the radius of the ring. For the ring to contribute $\sim$1\% of the measured $J_2R^2$ of our HST only fit, the ring would need to have a mass of $\sim10^{18}$ kg, about the mass of Namaka, given the known ring radius of 2287 km. For it to be the cause of Haumea's unexpectedly high \jt in the HST+Keck fit, the ring would need to be two orders of magnitude more massive, equivalent to tens of Hi'iaka masses. While no mass constraints on the ring are found in the literature, this value is absurdly high. \trackchange{For comparison, Saturn's rings are $\sim$1 Hi'iaka mass \citep{iess2019measurement}}. Hence, the ring is unlikely to contribute significantly to our measured \jt. There is a distinct possibility that more rings may be detected in future occultations \citep[e.g.][]{pereira2023two}, but even when combined, a ring system is unlikely to contain enough mass to substantially contribute to Haumea's \jt. 

Another potential source is an undetected satellite within Namaka's orbit. Averaged over an orbit, the putative inner satellite would act similar to a solid ring of material. Hence, using Equation \ref{eqn:ringj2} above to calculate the \jt{} of a putative inner satelite orbiting at 10000 km, we find that an inner satellite with a mass near Namaka's mass could significantly contribute to Haumea's measured \jt. Unfortunately, \citet{burkhart2016deep} significantly ruled out satellites more than 60 km in diameter closer than 10000 km by using a nonlinear shift-and-stack image analysis technique. This diameter implies a mass approximately one-tenth of Namaka's mass, which would scarcely contribute to the overall \jt. Given this, we believe it is unlikely that our results could be caused by an unknown inner satellite. Likewise, undiscovered satellites external to Hi'iaka's orbit would not produce the observed signature.

An alternative reason could be an extreme mass anomaly on Haumea's surface, either positive or negative, which would cause Haumea's \jt{} to be larger than expected. Unfortunately, the mass surplus (or deficit) would have to be substantial, of order $\sim$10\% of Haumea's total mass. Maintaining a mass anomaly of that magnitude would require Haumea to have implausibly high material strength. Using the method developed by \citet{johnson1973topography}, we optimistically estimate that Haumea may support a maximum topographic feature of $\sim$10 km, amounting to far less than the $\sim$100 km required to produce an unusually high \jt. Hence, it seems unlikely that a surface feature could plausibly explain the HST+Keck measurement.

Likewise, Haumea could have an unusual interior density distribution. If Haumea formed in the aftermath of a large collision, it may have an unusually shaped core left over as a remnant of this impact. Alternatively, its core could be offset relative to its external figure, potentially explaining Haumea's ``Dark Red Spot'' \citep{lacerda2009time}. Assuming Haumea's core is triaxial and adopting the external figure of Haumea as measured by stellar occultation, we can calculate the \jt{} of Haumea with an arbitrary triaxial core shape with an offset. We find that for any realistic core shape or offset, Haumea's \jt{} can not increase by more than 50\%, still well below the constraint provided by the HST+Keck fit. In any case, the extreme versions of these hypotheses are unlikely to be geophysically viable as they ignore fluid-like relaxation of Haumea's core and mantle. We believe that an unusual interior is unlikely the cause of our results. 

Due to the implausibility of all of these solutions, we conclude that our result is due to factors dependent on our modeling techniques or data. One possible explanation is higher order non-Keplerian dynamics taking place within the system. Since our model only includes Haumea's \jt, other gravitational harmonics may be needed to fully model the system. To investigate the effect of \ct, we ran a \multimoon{} fit that included Haumea's \ct{} potential. This fit gave nearly identical results and found no constraint on \ct{}, indicating that the orbital dynamics of the system are not strongly coupled to \ct. Indeed, previous work exploring the effects of \ct{} found that it is only relevant when $P_{orb} \sim P_{spin}$, or near a low order SOR \citep{proudfoot2021prolate}. Beyond quadrupole dynamics, fourth-order, or hexadecapole, dynamics could contribute to orbital precession, but as previously discussed their $r^{-5}$ dependency ensures that their contributions are small. \trackchange{Taking the $J_4$ harmonic as an example, we find that the ratio of nodal precession \trackchange{of Namaka's orbit} caused by the $J_4$ and $J_2$ harmonics is $\dot{\Omega}_{J_4} / \dot{\Omega}_{J_2} \approx 0.003 \frac{J_4}{J_2}$. Apsidal precession has a similarly small strength. Typically, the $J_4$ harmonic is much smaller than $J_2$, implying the nodal precession induced by $J_4$ is a very small effect.}

Odd harmonics (e.g., $J_3$, $C_{31}$, etc.) could, in theory, also play a role in the system's orbital dynamics. (Note that the ``dipole'' term is 0 due to using the center-of-mass as the coordinate system; a center-of-mass--center-of-figure offset can contribute to \jt{} which was included in the calculations with the offset core above). \trackchange{Again, finding the ratio of nodal precession for Namaka, we find $\dot{\Omega}_{J_3} / \dot{\Omega}_{J_2} \approx 0.006 \frac{J_3}{J_2}$ when $\dot{\Omega}_{J_3}$ is at its maximum. Odd harmonics are typically extremely small in bodies at (or near) hydrostatic equilibrium, producing $\sim$no detectable precession.} While it seems unlikely that $J_3$ or other odd harmonics cause significant changes in the dynamics on the timescale of our observational data, future investigations should explicitly test whether odd harmonics are necessary for accurate modeling of the Haumea system. 

\begin{figure}
    \includegraphics[width=\linewidth]{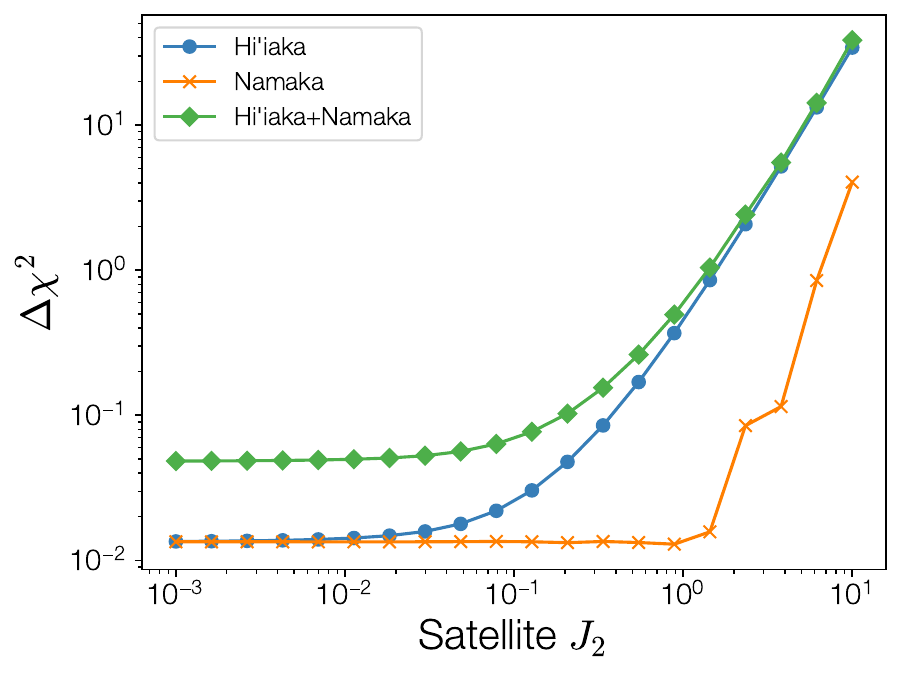}
    \caption{Change in orbit fit quality due to the nonspherical shapes of Haumea's satellites. In this plot, $\Delta \chi^2 = \sum^{i}(x_{i,J_2} - x_{i,J_2 = 0})/\sigma_{i,obs}$, where $x_{i,J_2}$ and $x_{i,J_2 = 0}$ are synthetic astrometric measurements from models including satellite \jt{} and those without. $\sigma_{i,obs}$ are the true measurement uncertainties for the system as tabulated in Table \ref{tab:observations}. This formulation allows us to directly determine whether the satellites' \jt values produce detectable changes in the system relative astrometry, given our current observations (HST+Keck). The parameters for the models used were taken from the best fit of our orbit fits. The rotational poles of the satellites were chosen to have high obliquities (w.r.t.~their Haumea-centric orbits) to enhance the effect of \jt. The three separate lines show each satellite's individual contribution, as well as a comparison where both satellites have (the same) \jt. For reasonable values, $J_2 \lesssim 0.5$, very little change in fit quality is found, although for extremely nonspherical shapes, Hi'iaka's \jt{} could begin to alter the fit.}
    \label{fig:satj2}
\end{figure}

Another possible effect that we do not account for is the satellites' putatively nonspherical gravitational fields. Our model assumes that Hi'iaka and Namaka are point masses, although they are likely to be substantially nonspherical. In some cases, however, the gravitational harmonics of a system's secondary can play a major role in the overall orbital dynamics \citep[e.g.][]{ragozzine2009probing}. \multimoon{} is well poised to explicitly test this assumption. Rather than adding six parameters to our overall model, which would be computationally expensive, we can simply add the satellites' harmonics as fixed values. We can then compare a model without their harmonics to one with them, allowing us to see the change in system dynamics. In this comparison, we use the characteristics of all our observations (HST+Keck) to explicitly connect the dynamical change to actual observability. In Figure \ref{fig:satj2}, we show the change in orbit fit quality as a function of Hi'iaka and Namaka's \jt, when both satellites have a moderately high obliquity. We define change in fit quality as $\Delta \chi^2 = \sum^{i}(x_{i,J_2} - x_{i,J_2 = 0})/\sigma_{i,obs}$, where $x_{i,J_2}$ and $x_{i,J_2 = 0}$ are synthetic astrometric measurements from models including satellite \jt{} and those without and $\sigma_{i,obs}$ are the measurement uncertainties for the system as tabulated in Table \ref{tab:observations}. Using those measurement uncertainties allows us to connect the system's dynamics to the actual observations. Overall, we find the fit quality is barely decreased when reasonable values for \jt{} are tested. Hi'iaka would need $J_2 > 2$ for a detectable change, while Namaka would need $J_2 > 10$. For comparison, Arrokoth, a contact binary, has a $J_2 \sim 0.3$. While not an exhaustive search of parameter space, this is strong evidence that Hi'iaka and Namaka's nonspherical shapes do not significantly contribute to the system's dynamics with the present data. 

In our view, the only remaining option is the presence of unknown systematic errors plaguing our dataset. Despite our use of novel statistical techniques, our model cannot account for all systematic errors arising from combining our dataset. For example, time-varying distortions in the NIRC2 field cannot be appropriately accounted for by our model. Likewise, wavelength-dependent offsets between Haumea's center of light and center of mass, potentially caused by the known wavelength-dependent rotational variability known as the ``Dark Red Spot'' \citep{lacerda2009time} may introduce unwanted systematics when combining the data sets. Indeed, when combining the datasets, we find that the Keck data from the 2020s is incompatible with the HST visit from 2014. Our combined dataset produces large residuals for the 2014 HST visit, while the HST only fit shows no such effect. While disconcerting, this conclusion is not extremely surprising given a similar result in RB09, from which we draw much of our data. Those authors similarly found that the Keck data was inconsistent with the HST-only fits. We thus argue that unknown systematic errors are the source of our unusually high measurement of \jt. HST instruments are extremely well-studied and have been rigorously cross-validated, so we view the HST only fit as more trustworthy. To remain as conservative as possible, we adopt the HST only fit as our nominal model for the rest of the analysis in this work. 

\subsection{Masses and Densities of Hi'iaka and Namaka}
\label{sec:dis_mass}
While our model does not fully constrain the masses of the satellites, especially Hi'iaka, it is still scientifically valuable given some assumptions. If we assume Haumea has \jt{} of a differentiated body ($J_2 = 0.16$ using the \citet{dunham2019haumea} model), we are able to estimate the mass of Hi'iaka to be roughly $1.6 \times 10^{19}$, with uncertainties of $\sim$10\%. This is similar to the mass determination found in RB09, albeit 10\% smaller. Given an estimated radius of $\sim$160 km (RB09), this gives a density of $\sim$900 kg m$^{-3}$, similar to other TNOs in the same size range \citep{grundy2019uk126}. Even given the undifferentiated model, Hi'iaka is not substantially less massive, altering the density marginally. 

While the mass measurement for Hi'iaka must be improved to place better constraints on Hi'iaka's density, this needs to be matched with improvement in the radius determination. As density is proportional to $r^{-3}$ (as opposed to linear in mass), uncertainties in radius have a stronger influence on the overall density uncertainty. Thermal observations and modeling of the satellites indicate high albedos and small sizes \citep{muller2019haumea}, but uncertainties are large. At this time, given the massive uncertainties in both radius and mass, robust interpretation of the density of Hi'iaka is impossible. 

Fortuitously, however, a recent occultation campaign has captured a multi-chord stellar occultation of Hi'iaka \trackchange{(Fernandez-Valenzeula et al., in prep)}. The results of this campaign are forthcoming, but early indications point towards Hi'iaka being larger than expected \trackchange{(E. Fernandez-Valenzeula, pers. comm.)}. When using the results from those observations, future work should be able to tightly constrain the density of Hi'iaka if the Hi'iaka mass--\jt{} degeneracy is broken. 

While our uncertainty on the mass of Namaka is much smaller than for Hi'iaka, Namaka still suffers from a poorly constrained size. However, given size estimates from thermal measurements \citep{muller2019haumea} and past photometric analysis (RB09), Namaka could have a radius of $\sim$75 km, giving a density of $\sim$650 kg m$^{-3}$. While quite low, it is typical of small TNOs \citep{grundy2019uk126}. Such a low density is indicative of a porous interior, consistent with relatively gentle accretion in a satellite forming disk around Haumea in the aftermath of an impact. To improve confidence in this measurement, Namaka should be a high-priority target for an occultation campaign. 

\subsection{Haumea's Pole}
\label{sec:dis_pole}
Among TNOs, very few spin poles have been constrained. When disregarding non-Keplerian fitting, the only techniques currently able to characterize the spin poles of TNOs are long-term light curve monitoring and occultations. Light curve inversion techniques require observations of a TNO over a significant portion of their orbit, which is \trackchange{difficult} due to TNOs' long heliocentric orbital periods. Occultations are extremely powerful for inferring spin poles, but observations of multiple multi-chord events are required, which are only available for a few TNOs. Non-Keplerian orbit fitting now adds an additional tool which can be used to understand the spin poles of TNOs\footnote{Technically, non-Keplerian fitting finds the pole direction of the nonspherical gravitational field, not the figure of the overall body. However, in practice these are functionally identical, especially for large objects like Haumea (Ragozzine et al., submitted).}. 
Normally, non-Keplerian fitting cannot find a unique pole solution, but is able to determine the angle between the primary's spin pole and the secondary's orbit normal. However, since Haumea has two satellites, the spin pole can be found unambiguously. 

Our measurement in this work represents the first dynamically determined spin pole among TNOs. We are able to place tight constraints on the spin pole direction of Haumea, finding $(\alpha_p, \delta_p) = (282.9^{+0.6}_{-0.7} \degr,-9.7^{+0.6}_{-1.0} \degr)$. \trackchange{Interestingly, Haumea's pole is almost perpendicular to its orbit. Based on Haumea's heliocentric orbit, we find Haumea's obliquity to be 87.1$\degr$.} Our dynamically determined spin pole measurement lies 2.3$\degr$ from the ring pole determined by stellar occultation $(\alpha_p, \delta_p) = (285.1\pm0.5 \degr,-10.6\pm1.2 \degr)$ \citep{ortiz2017size}. Given that our methods are completely different from those used in analyzing the stellar occultation and we use no constraints from that work, the close match is encouraging. Formally, however, these two spin pole measurements are $\sim$2$\sigma$ apart. The difference could be a real effect (i.e., Haumea's ring is inclined with respect to Haumea's nonspherical gravitational field), but the nodal precession of ring particles induced in this case is likely to erode the rings excessively \citep{marzari2020ring}. It seems much more likely that the ring is coplanar with Haumea's gravitational field, and the disagreement is due to model dependent factors or random chance. For example, \citet{ortiz2017size} modeled Haumea's ring as a flat, circular annulus, but the true ring may have substantial eccentricity and/or thickness. Alternatively, since the spin pole in our fit is highly correlated with \jt, our large uncertainties may cause/contribute to the disagreement. When examining our HST+Keck fit, we find the spin pole measurements are even further apart, lending further credibility to the HST only model.

\begin{figure*}
    \includegraphics[width=\textwidth]{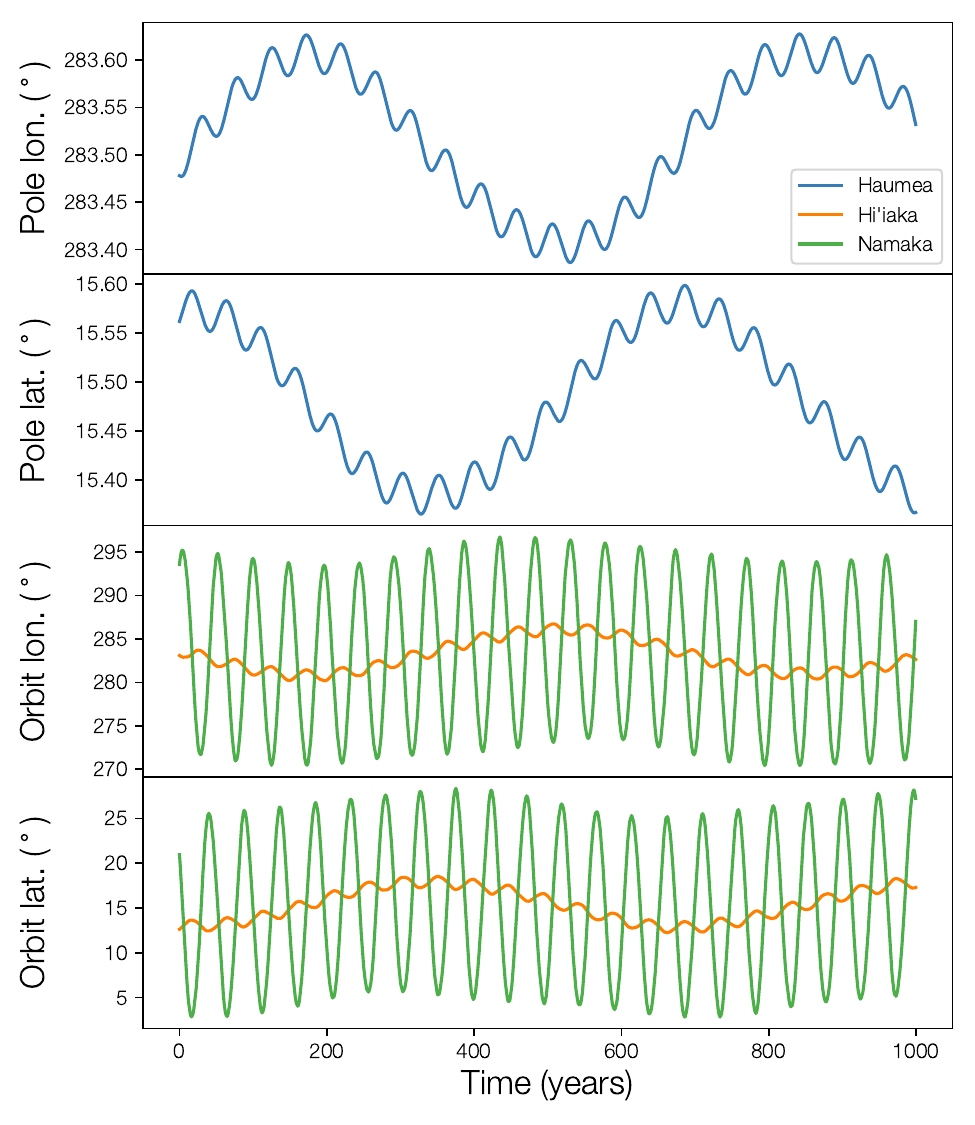}
    \caption{The spin precession of Haumea over a 1000 year integration. In the top two panels, we show the spin precession of Haumea in terms of pole ecliptic longitude and latitude. In the bottom two panels we show the precession of the orbit normal of Hi'iaka and Namaka, again in terms of ecliptic longitude and latitude. The integration parameters have been chosen to be representative of the posterior found in Table \ref{tab:fits}. Similar integrations with different values for Haumea's \jt{} change the long-term precession period, but are qualitatively similar. Haumea's spin precession is coupled with the nodal precession of the satellites. High frequency, low amplitude variations in Haumea's pole direction are caused by Namaka's rapid precession, while low frequency, high amplitude variations are coupled with Hi'iaka's \jt{} precession. As can be seen, the precession of all components are strongly coupled, both through Hi'iaka-Namaka gravitational interactions and interactions with Haumea's \jt. }
    \label{fig:haumea_precession}
\end{figure*}

\subsection{Rotational Dynamics} \label{sec:spindynamics}
The rotational dynamics of the Haumea system are extremely interesting to study. For Haumea itself, the torque from both satellites on Haumea's nonspherical body cause a small amount of axial precession. We show a 1000 year integration of Haumea's spin dynamics in Figure \ref{fig:haumea_precession}. The integration is performed by \texttt{SPINNY}, the spin-orbit integrator at the heart of \multimoon{} (Ragozzine et al., submitted). We find that Haumea's axis precesses by $<1\degr$ on a timescale of 100s of years. Visible is a complex precession cycle in Haumea's spin pole direction with one `fast' frequency and one `slow' frequency. These two frequencies are caused by the torque from each satellite with periods corresponding to each satellite's nodal precession period. Namaka's nodal precession period is strongly coupled with both Hi'iaka-Namaka interactions and Namaka-$J_2$ interactions. Since Namaka is coupled to Hi'iaka's nodal precession, Namaka's nodal precession then weakly couples to Hi'iaka-$J_2$ nodal precession, although this is a much smaller effect. Hi'iaka's nodal precession has a fast, low-amplitude component caused by Hi'iaka-Namaka interactions as well as a slow, high-amplitude component from the Hi'iaka-$J_2$ interaction. The low-amplitude, high frequency precession of Haumea's pole caused by Namaka would produce little detectable change in Haumea's light curve or occultation shadow. The Hi'iaka coupled precession has a much higher amplitude, but has a period of 100s of years, severely hampering detectability. Given Haumea's prolate shape, the satellites' torques can also alter Haumea's rotation period, however, this effect is tiny due to Haumea's large angular momentum. Using \texttt{SPINNY} simulations, we estimate the period variations are $\sim10^{-8}$ hours, approximately two orders of magnitude smaller than the uncertainty in the measured rotation period \citep{rabinowitz2006photometric}.

\begin{figure*}
    \includegraphics[width=\textwidth]{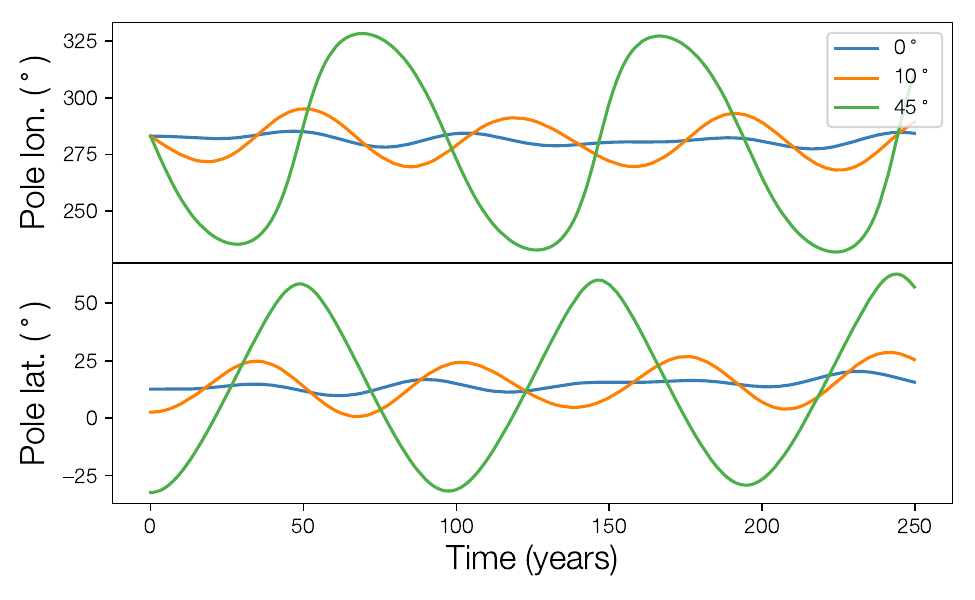}
    \caption{The precession of Hi'iaka's spin axis based on its initial obliquity. Similar to Figure \ref{fig:haumea_precession}, we show the precession of Hi'iaka's spin axis in terms of ecliptic longitude and latitude. Initial integration parameters have been chosen to be representative of the posterior found in Table \ref{tab:fits}. The moments of inertia of Hi'iaka were chosen to be similar to objects of similar size, although their values only change the frequency of the precession. For different initial obliquities, we find different precession periods, matching analytical theory and results in the literature \citep[e.g.][]{hastings2016short}. Interestingly, there are small variations when Hi'iaka's spin is initially aligned with its orbit. Although initially there would be no net torque and no precession when aligned, since Hi'iaka's orbit precesses, the alignement is broken and precession begins.}
    \label{fig:hiiaka_precession}
\end{figure*}

\begin{figure*}
    \includegraphics[width=\textwidth]{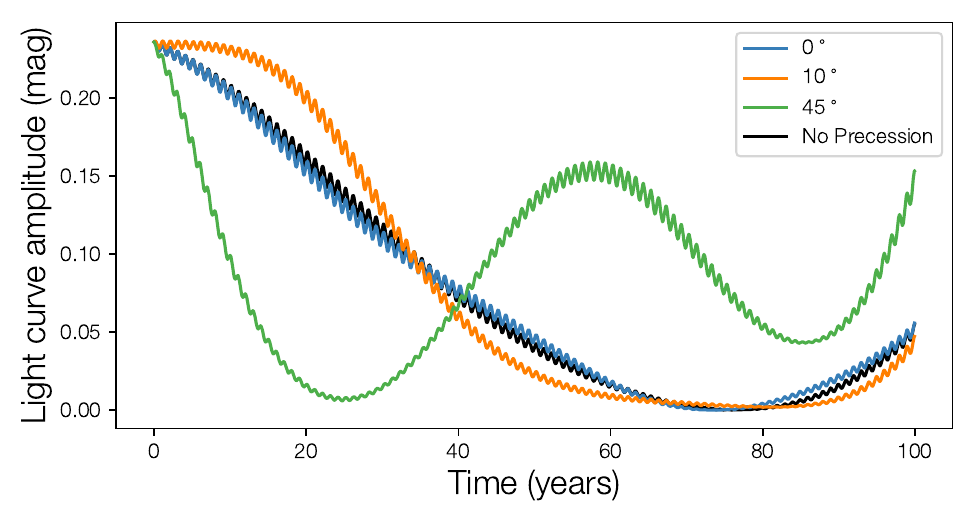}
    \caption{The change in Hi'iaka's light curve amplitude over time. Using the integrations similar to those shown in Figure \ref{fig:hiiaka_precession}, we calculate the light curve amplitude using Equation 1 from \citet{hastings2016short}. Note that the x-axis is different than Figure \ref{fig:hiiaka_precession}. Axis ratios were chosen to be similar to other solar system objects at similar size, as well as approximately matching the light curve amplitude found in the literature \citep{hastings2016short}. The no precession case is found by taking a fixed pole direction. The fast variations in these functions are due to Earth's heliocentric orbit. Even for small obliquities, the light curve evolves significantly differently than the no precession case. As in Figure \ref{fig:hiiaka_precession}, we find that the aligned case (0$\degr$) still shows precession and slightly different light curve evolution when compared to the no precession case. Long-term monitoring of Hi'iaka's light curve may permit direct measurement of its spin pole over decadal timescales. 
    }
    \label{fig:hiiaka_lc}
\end{figure*}

More amenable to detectability is possible precession of Hi'iaka's rotational axis. In \citet{hastings2016short}, the light curve of Hi'iaka was studied using resolved photometry from HST images. They found that Hi'iaka is rapidly rotating ($\sim$9.8 hour double-peaked period) with an unusual sawtooth shaped light curve of amplitude $\Delta m \approx 0.23$. Using a simplified model of axial precession, they found that Hi'iaka's axial precession would be detectable on decade timescales if there was significant obliquity (w.r.t.~its orbit). The detectability is significantly enhanced by Hi'iaka's high amplitude light curve. Detection of any precession would require long-term monitoring of Hi'iaka's light curve which would slowly change amplitude and/or shape across the precession cycle. \texttt{SPINNY} provides an ideal framework for validating this possible method. Using the best fit from our nominal orbit model, we have explicitly modeled Hi'iaka's axial precession. In Figure \ref{fig:hiiaka_precession}, we show the evolution of Hi'iaka's rotation axis assuming differing starting obliquities. Then in Figure \ref{fig:hiiaka_lc}, we illustrate how that precession translates to variation in Hi'iaka's light curve amplitude assuming triaxial shapes as in \citet{hastings2016short}. Interestingly, even in the case where Hi'iaka's pole is initially aligned with its orbit, precession still occurs. While initially surprising, the precession is due to Hi'iaka's nodal precession in its orbit around Haumea which will always misalign Hi'iaka's spin pole. Encouragingly, even for a relatively small obliquity of $10\degr$, the precession is substantially different from the no precession case. This allows us to confirm previous results \citep[e.g.][]{hastings2016short} and show that small perturbations (e.g., Hi'iaka's eccentric orbit, torques from Namaka, etc.) seem to make little difference to the overall evolution. In the future, \texttt{SPINNY} and/or \multimoon{} could be modified to explicitly model changes in light curve amplitudes. This method would provide a detailed model with which to understand the spin dynamics of Hi'iaka; we defer this to future work.
\begin{figure*}
    \includegraphics[width=\textwidth]{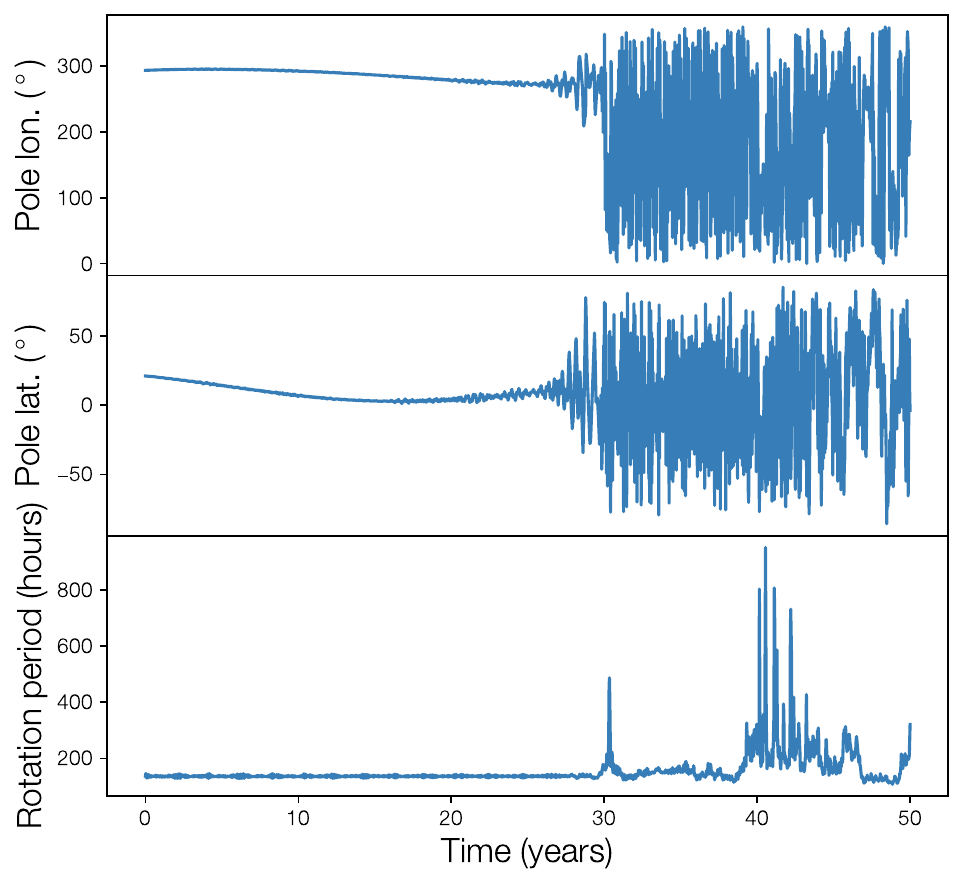}
    \caption{Chaotic rotation of Namaka. When initialized, Namaka rotates once every 6 days and is aligned with its orbit. After just a few decades, Namaka becomes attitude unstable and begins to tumble. Chaos is a natural consequence of Namaka's eccentric orbit and is inevitable if Namaka has been substantially despun. In this sense, Namaka is very similar to Saturn's satellite Hyperion, which chaotically rotates due to its eccentric orbit around Saturn \citep{wisdom1984chaotic,klavetter1989rotation}.}
    \label{fig:namaka_chaos}
\end{figure*}

Even though Namaka has been solidly detected in several epochs of HST observations, no periodic brightness variations have been found, although photometry from HST programs is suggestive of a long rotation period ($>$1 day). Purely based on theoretical dynamical arguments, Namaka is expected to be significantly despun, except if its initial rotation period was extremely short \citep{hastings2016short}. Given Namaka's eccentric orbit, overlap between SORs inevitably causes spin-orbit chaos, very similar to Saturn's satellite Hyperion \citep{wisdom1984chaotic}. As an illustration of chaotic rotation, the resonance overlap criterion, which predicts chaotic spin-orbit evolution if satisfied, near the 1:1 and 3:2 SORs is:

\begin{align}
    \sqrt{\frac{3(\mathcal{B}-\mathcal{A})}{\mathcal{C}}} \geq \frac{1}{2+\sqrt{14e}}
\end{align}
\noindent where $\mathcal{A}$, $\mathcal{B}$, and $\mathcal{C}$ are Namaka's principal moments of inertia and $e$ is Namaka's orbital eccentricity. In this case, to avoid spin-orbit chaos Namaka's shape would have to satisfy $\frac{(\mathcal{B}-\mathcal{A})}{\mathcal{C}} < 0.025$. Satellites that are similar in size to Namaka generally have $\frac{(\mathcal{B}-\mathcal{A})}{\mathcal{C}} \gtrsim 0.1$. For example, with a mass between Namaka and Hi'iaka ($\sim$5$\times 10^{18}$ kg), Hyperion has $\frac{(\mathcal{B}-\mathcal{A})}{\mathcal{C}} \approx 0.26$.

While the above calculation is a simplistic example comparing just two resonances, in general, resonance overlap and spin-orbit chaos are expected for a slowly rotating Namaka. In Figure \ref{fig:namaka_chaos}, we show the initiation of chaotic tumbling where Namaka's attitude (i.e. pole direction) and rotation period evolve chaotically. Given the difficulty of acquiring resolved photometric observations of Namaka and the long timespan needed to detect a slow (and possibly chaotic) rotation, confirming it may be extremely difficult. Similarly, searching for Namaka's light curve in unresolved photometry of the entire system is extremely difficult as Namaka's brightness only contributes $\sim$1\% of the total system flux. 

\subsection{Ring-Satellite Interactions}
Haumea's ring, first discovered during a stellar occultation, lies close to Haumea's equatorial plane, as shown in the previous section. This matches theoretical expectations, which show that ring particles should collisionally damp to the equatorial plane in the presence of Haumea's \jt. When accounting for interactions between ring particles and Haumea's satellites, however, ring particles can preferentially settle in the satellite orbit plane \citep{marzari2020ring}. However, the satellites are too far away and/or not massive enough to cause this to occur for the ring. The satellites' main dynamical roles are to act as small perturbers that increase velocity dispersion. \citet{marzari2020ring} studied the increase in collision velocity of ring particles under the influence of Haumea's satellites and found that collision velocities still remained low with typical velocities $<1$ m s$^{-1}$. 

It may be possible that if other rings exist external to the currently known ring, satellite-ring dynamics may be more important. As the strength of satellite-ring interactions increases, the collision velocities between ring particles may become large enough to completely disperse the ring. Rings external to the Roche limit may be possible at the temperatures of the Kuiper belt \citep{morgado2023dense,pereira2023two}, but in Haumea's case they are likely to be in a regime where perturbations make them long-term unstable. 

\begin{deluxetable*}{ccCCCCCCCCCCCC}
\tablecaption{System Ephemeris}
\tablehead{
Julian Date & Date & \Delta x_N & \Delta y_N & \sigma_{\Delta x_N} & \sigma_{\Delta y_N} & $r_N$ & $\sigma_{r_N}$ & \Delta x_H & \Delta y_H & \sigma_{\Delta x_H} & \sigma_{\Delta y_H} & $r_H$ & $\sigma_{r_H}$ \\ 
& & \textrm{('')} & \textrm{('')} & \textrm{('')} & \textrm{('')} & \textrm{('')} & \textrm{('')} & \textrm{('')} & \textrm{('')} & \textrm{('')} & \textrm{('')} & \textrm{('')} & \textrm{('')}
}
\startdata
2453371.500 & 2005-01-01 00:00:00 & 0.068 & -0.190 & 0.002 & 0.005 & 0.202 & 0.004 & 0.057 & 0.649 & 0.001 & 0.003 & 0.651 & 0.003 \\
2453371.833 & 2005-01-01 08:00:00 & 0.068 & -0.121 & 0.002 & 0.005 & 0.139 & 0.004 & 0.043 & 0.600 & 0.001 & 0.003 & 0.601 & 0.003 \\
2453372.167 & 2005-01-01 16:00:00 & 0.068 & -0.051 & 0.002 & 0.005 & 0.085 & 0.003 & 0.029 & 0.549 & 0.001 & 0.003 & 0.550 & 0.003 \\
2453372.500 & 2005-01-02 00:00:00 & 0.067 & 0.019 & 0.002 & 0.005 & 0.069 & 0.002 & 0.015 & 0.497 & 0.001 & 0.003 & 0.497 & 0.003 \\
2453372.833 & 2005-01-02 08:00:00 & 0.065 & 0.090 & 0.002 & 0.006 & 0.111 & 0.004 & 0.000 & 0.444 & 0.001 & 0.003 & 0.444 & 0.003 \\
2453373.167 & 2005-01-02 16:00:00 & 0.062 & 0.159 & 0.002 & 0.006 & 0.171 & 0.005 & -0.014 & 0.391 & 0.001 & 0.003 & 0.391 & 0.003 \\
2453373.500 & 2005-01-03 00:00:00 & 0.059 & 0.227 & 0.002 & 0.006 & 0.234 & 0.005 & -0.028 & 0.336 & 0.001 & 0.003 & 0.337 & 0.003 \\
2453373.833 & 2005-01-03 08:00:00 & 0.055 & 0.291 & 0.002 & 0.006 & 0.296 & 0.005 & -0.042 & 0.281 & 0.001 & 0.003 & 0.284 & 0.003 \\
2453374.167 & 2005-01-03 16:00:00 & 0.051 & 0.352 & 0.002 & 0.005 & 0.355 & 0.005 & -0.056 & 0.225 & 0.001 & 0.003 & 0.232 & 0.003 \\
2453374.500 & 2005-01-04 00:00:00 & 0.045 & 0.407 & 0.002 & 0.005 & 0.410 & 0.005 & -0.070 & 0.168 & 0.001 & 0.004 & 0.182 & 0.003 \\ 
\hline
\enddata
\tablecomments{The predicted right ascension and declination positions of Namaka (N) and Hi'iaka (H) from 2005 to 2035. $\Delta x$ and $\Delta y$ are the predicted right ascension and declination, $r$ is the total separation, and $\sigma$ are the uncertainties on each value. All values are given in arcseconds. Predicted positions, separations, and uncertainties are taken from a sample of 500 random posterior draws. We display the first 10 rows of the table with the rest of the table available as a machine-readable table. }
\label{tab:ephem}
\end{deluxetable*}

\subsection{Future and past observations}
To enable future observations of the Haumea system, we have created an ephemeris of the system containing the predicted $\Delta \alpha \cos{\delta}$ and $\Delta \delta$ of each satellite and the uncertainties on the positions. We compute the ephemeris using 500 random draws from our posterior distribution. The ephemeris contains the predicted positions every 8 hours between 2005 and 2035. In Table \ref{tab:ephem}, we show the first 10 lines of the ephemeris. The ephemeris is published in its entirety in machine-readable format. Uncertainties on our predictions for both satellites are quite small ($\lesssim$50 mas) through the 2020s until 2030, at which time, the uncertainties begin to grow rapidly, especially for Namaka. By 2035, uncertainties on Namaka's position are of order $\sim$0.1 arcseconds. Rapid growth in uncertainty is attributable to the large degeneracy in our model (see Table \ref{tab:fits} and Figure \ref{fig:degneracy}). 

To ascertain whether future observations may be able to break the mass-\jt{} degeneracy in our fits, we have analyzed an ephemeris (similar to that presented above) where the predicted positions are also a function of Haumea's \jt{} (or equivalently Hi'iaka's mass). We find that the future positions of Hi'iaka and Namaka are strong functions of Haumea's \jt{} (or Hi'iaka's mass) indicating that the degeneracy will be broken with additional HST observations. Based on our fits, we can roughly estimate that $\sim$2-4 new epochs of observations in the mid-2020s (or beyond) are necessary to constrain Haumea's \jt{} with $\sim$10\% precision. Past work has shown that timing observations at certain, well-selected times can substantially improve the quality of the resulting orbit fits (Proudfoot et al., submitted). These times occur when the uncertainties in $\Delta \alpha \cos{\delta}$ and $\Delta \delta$ in Table \ref{tab:ephem} are at (local) maximum. We recommend that continued observations be taken by HST to prevent any systematic errors from arising, as we found in our HST+Keck fits. As these observations will probe Haumea's interior and place strong constraints on its differentiation, we regard them as high priority.

As well as its use for future observations, our ephemeris can evaluate past predictions of mutual events in light of our new orbit solution. We find that the mutual event predictions made in RB09 are generally accurate, even given their sign error in the system astrometry. We predict events that are similar in depth and duration, but are somewhat offset in time. The timing differences are only a few hours in 2009-2011, but steadily grow to tens of hours by the end of the mutual event season. Since the mutual event timings are quite sensitive to Namaka's eccentricity, and we find an eccentricity $\sim$15\% lower than previously found, it is unsurprising that we find timing differences. The next mutual event season will occur in approximately half a heliocentric Haumea orbit, about midway through the 2200s. The exact timeframe will be dependent on Namaka and Hi'iaka's precise precession rates which future observations will be able to precisely measure.

\subsection{The Origin and Evolution of Haumea's Satellites}

Our new information on the properties of the Haumea system give more accurate insight into the formation and evolution of Haumea and its satellites. 

Presumably, Haumea's satellites (and possibly its ring) formed in the same event that created the Haumea family. \trackchange{For a full review of family formation proposals, see \citet{proudfoot2019modeling}. The most compelling proposals involve mass shedding due to an excess of angular momentum, explaining Haumea's rapid rotation and the family members' low ejection velocity. There is still much} debate in the literature on the original reason that proto-Haumea had too much angular momentum: \citet{proudfoot2022formation} propose the merger of a near-equal-mass binary similar to Pluto-Charon perhaps triggered by Kozai cycles initiated due to Haumea's placement on a higher inclination orbit. \citet{noviello2022let} point out that internal evolution could change \trackchange{Haumea's moment of inertia, spinning it up past the point of break-up}. \citet{ortiz2012rotational} and others suggest that it may be the cumulative effect of many smaller impacts, though starting with a rapid rotation would significant increase the probability that a random walk would lead to such a rapid rotation. 

Whatever the origin, the general agreement is that Haumea goes through a phase of excess angular momentum where water ice chunks are ejected at low velocities (relative to catastrophic collisions) from the tips of Haumea. Thus, a plausible starting point for the formation of the satellites is a disk of satellitesimals ejected from the fast-spinning Haumea, mostly composed of pure water ice with sizes similar to known satellites and family members. Given that the distribution of ejection (or sub-ejection) velocities should be somewhat smooth, it is likely that satellitesimals are both ejected and remain in orbit.

This initial disk of satellitesimals should rapidly evolve through ejections and collisions. In this scenario, the unejected material experiences a collisional cascade that leads to a final configuration of a ring of near-circular, near-coplanar disk of collisional debris. This debris would eventually reaccrete into Hi'iaka and Namaka. Haumea's ring could also derive from the parts of this initial ring that did not form into satellites. \trackchange{Combining smoothed particle hydrodynamic (SPH) modeling with long-term dynamical evolution is necessary to understand this process}. We strongly encourage work on this problem, which may provide understanding applicable to the formation of other TNO satellites. 

After their initial formation, a variety of physical processes can influence the evolution of the satellites until they reach their current configuration. The most important effects are expected to be tidal evolution, Hi'iaka-Namaka interactions, excitation from passing TNOs and binaries, and possible interactions from previous satellites \citep{cuk2013dynamics, hastings2016short}. These are discussed in detail in \citet{cuk2013dynamics} and we focus here only things that are updated in our new fit. 

\citet{cuk2013dynamics} found that long-term orbital stability would be significantly improved if the satellites were $\sim$50\% of their nominal masses reported in \citet{rb09}. Indeed, our new results are most consistent with satellites that are $\sim$60\% of the initially estimated masses with the Namaka/Haumea mass ratio of 3.0 $\pm$ 0.6 $\times$ 10$^{-4}$ and the Hi'iaka/Haumea mass ratio of 3.1 $\pm$ 0.8 $\times$ 10$^{-3}$ in the HST-only fit (see Table \ref{tab:fits}), though this is strongly affected by the degeneracy discussed above. 

\citet{cuk2013dynamics} also investigate in detail the effect of the 8:3 and 3:1 mean-motion resonances between the satellites, especially in the presence of tidal evolution. This was based on the observation by \citet{rb09} that the satellites were possibly in or near the 8:3 resonance, suggesting that tidal evolution in resonances could explain the source of the moderately eccentric and inclined orbits. This idea was further strengthened by the observation -- still true with the new orbit fit -- that the excitation is similar in eccentricity and inclination and is inversely proportional to the masses of the satellites. This means that the ``Angular Momentum Deficit'' \citep{Laskar1997large} is approximately evenly partitioned between the two satellites, suggesting they have been strongly dynamically coupled at some point in the past (or present). We find a period ratio of Hi'iaka and Namaka of 2.689 $\pm$ 0.004 (including corrections to Newton's Version of Kepler's Third Law from $J_2$), slightly closer to the 8:3 mean-motion resonance that reported in \citet{rb09}. With the residual uncertainty in Hi'iaka's mass and Haumea's $J_2$, we leave to future investigation whether the 8:3 resonance is currently dynamically active in the system. Confirmation of an active resonance is key to understanding the recent history of the satellites. Overall, however, the general conclusion of \citep{cuk2013dynamics} that resonance passage could explain the excited orbits remains consistent with the updated fits. 

The primary challenge in explaining the current orbital configuration of the satellites is their distant orbits from Haumea at $\sim$36 and $\sim$70 primary radii. Tidal evolution to such distances is challenging in standard tidal theories, requiring Haumea tides to be extremely dissipative with an implausible combination of tidal parameters. This is exacerbated by a factor of $\sim$2 with the lower masses for the satellites, but \citet{quillen2016tidal} find that the triaxial shape of Haumea increases tidal evolution by a factor of a few (though not as large as hoped for by \citet{rb09} and \citet{cuk2013dynamics}). \trackchange{It is possible that more realistic tidal and geophysical models are able to resolve these issues. It is interesting to note that the satellites are close to the expected outcome of tidal evolution given their mass and semi-major axis ratios.}

One potential resolution to the tidal dissipation problem is to form Hi'iaka and Namaka near their current locations. Since tidal expansion is very strongly dependent on separation, even starting at $\sim$90\% of the present distance does not relieve pressure on tidal theories. \trackchange{Given that we observe objects ejected from this disk, it could have been extended out to the Hill sphere. A disk of such size may allow satellite formation further out than previously thought}. Along these lines, we note that although the satellites seem well-separated, dynamically speaking they are only separated about 5 mutual Hill radii, suggesting they are dynamically packed. Hence, intermediate satellites could not fit dynamically, so perhaps Namaka and Hi'iaka are the natural outcome of an extended disk near their present locations. \trackchange{Further modeling is needed to understand the plausibility of this scenario.}

In conclusion, the formation and evolution of Namaka and Hi'iaka are plausibly connected to the same process that spun up Haumea and created the family. One possible formation hypothesis is that water ice chunks which do not escape to form the family eventually collide and grind down to a disk near the present location of the satellites. Namaka and Hi'iaka perhaps form directly from this disk and recent dynamical interactions (e.g., from the nearby 8:3 resonance) lead to the orbits seen today, as proposed in \citet{cuk2013dynamics}. Once Hi'iaka's mass is better known, a more detailed investigation into the secular, resonant, and tidal dynamics could confirm or refute this hypothesis. However, the most important next step is more detailed modeling of the post-spin-up and family ejection process, extending to the longer timescals needed to understand the subsequent epoch of satellite formation. 

\section{Conclusion}
\label{sec:conclusions}
Using a state-of-the-art orbit fitter, \multimoon, combined with several new epochs of observations from Keck and HST, we have refit the orbits of Haumea's satellites. The model we use can account for both satellite-satellite interactions and Haumea's oblate gravitational field. We find that unaccounted systematic errors are present when fitting to the combined HST and Keck datasets, even when using robust statistical techniques that can account for some types of systematics. Although the HST+Keck fit can precisely constrain Haumea's \jt{} and the masses of Hi'iaka and Namaka, we reject this fit since it has unreasonably high residuals and predict physically unrealistic values for Haumea's \jt. On the other hand, our orbit fit to only the HST data provides a better fit to the data overall. Unfortunately, this fit suffers from a degeneracy between Hi'iaka's mass and Haumea's \jt, preventing a precise measurement of these two parameters. 

For our HST only orbit fit, we detect Haumea's \jt{} at $\sim2.5\sigma$ confidence ($J_2 = 0.262^{+0.103}_{-0.112}$). Our fits are unable to discriminate between either a homogeneous or differentiated interior, but only a few additional epochs of precise astrometric observations will easily provide the precision to distinguish between these models. Our fit has also provided a measurement of Haumea's rotational pole $(\alpha_p, \delta_p) = (282.9^{+0.6}_{-0.7} \degr,-9.7^{+0.6}_{-1.0} \degr)$, which lies extremely close to the orbit pole of Haumea's ring \citep{ortiz2017size}. In this sense, we presume that Haumea's ring lies in Haumea's equatorial plane and is minimally perturbed by Hi'iaka and Namaka. Determining Haumea's pole allows us to place tight constraints on the inclination of the satellites w.r.t.~Haumea's equator, showing that Hi'iaka and Namaka are inclined by approximately $1.01^{+0.66}_{-0.47}\degr$ and $12.79^{+1.01}_{-0.58}\degr$, respectively. Both Hi'iaka and Namaka are on somewhat excited orbits, shown in both their inclination and eccentricity, hinting at past dynamical excitation \citep{cuk2013dynamics}. 

Using our orbit fits, we have also characterized the rotational dynamics of the Haumea system using the spin-orbit integrator \texttt{SPINNY} (Ragozzine et al., submitted). We find that Haumea's rotation axis precesses $<0.5\degr$ on $\sim$kyr timescales, and is most strongly coupled to Hi'iaka's slow precession due to Haumea's \jt. Hi'iaka's rotation is expected to strongly precess on decadal timescales, which should have strong effects on the evolution of its light curve. Namaka is expected to rotate extremely slowly, based on both dynamical/tidal arguments and preliminary studies of its light curve. This putative slow rotation implies that Namaka chaotically rotates due to its significantly eccentric orbit. 

To enable future observations of the Haumea system we have generated a satellite ephemeris over the next decade. These observations will enable a probe of Haumea's interior, aid in understanding the spin states of Haumea's satellites, and continue to provide insights into Haumea's formation. Understanding the Haumea system as a whole is crucial for understanding large TNO formation and evolution. The production of satellites and satellite systems seems to be ubiqitous across the transneptunian region, but the processes at play are still not well-explored. Thankfully, continued observations of Haumea and its satellites will enable deeper knowledge of the far reaches of our solar system.

\begin{acknowledgments}
We thank Simon Porter and Seth Pincock for help with \texttt{SPINNY} and Steve Desch for valuable discussions on Haumea's origin and interior. We also thank the BYU Office of Research Computing for their dedication to providing computing resources without which this work would not have been possible. 

The authors wish to recognize and acknowledge the very significant cultural role and reverence that the summit of Maunakea has always had within the Native Hawaiian community. We are most fortunate to have the opportunity to conduct observations from this mountain. 

Some of the data presented herein were obtained at Keck Observatory, which is a private 501(c)3 non-profit organization operated as a scientific partnership among the California Institute of Technology, the University of California, and the National Aeronautics and Space Administration. The Observatory was made possible by the generous financial support of the W. M. Keck Foundation. These data were obtained from telescope time allocated to NASA through the agency's scientific partnership with the California Institute of Technology and the University of California. Acquisition of the data was supported by NASA Keck PI Data Awards, administered by the NASA Exoplanet Science Institute.

This research is based on observations made with the NASA/ESA Hubble Space Telescope obtained from the Space Telescope Science Institute, which is operated by the Association of Universities for Research in Astronomy, Inc., under NASA contract NAS 5–26555. These observations are associated with programs 10545, 10860, 11169, 11518, 12243, and 13873. Support for this work was funded by NASA through grants HST-GO-12243, HST-GO-13873, and HST-AR-14581. This work was also supported by grant 80NSSC19K0028 from NASA Solar System Workings. 

\end{acknowledgments}

\bibliographystyle{apj}
\bibliography{all}

\end{document}